\documentclass[prb,twocolumn,groupedaddress,amsmath,floatfix]{revtex4}
\usepackage{graphicx}
\usepackage{bm}

\newcommand{\brho}{{\bm{\rho}}}

\newcommand{\grad}{{\bm{\nabla}}}
\newcommand{\zh}{\hat{\bf z}}
\newcommand{\xh}{\hat{\bf x}}
\newcommand{\yh}{\hat{\bf y}}
\newcommand{\rh}{\hat{\bf r}}

\newcommand{\br}{{\bf r}}
\newcommand{\be}{\begin{equation}}
\newcommand{\ee}{\end{equation}}
\newcommand{\bea}{\begin{eqnarray}}
\newcommand{\eea}{\end{eqnarray}}
\newcommand{\bse}{\begin{subequations}}
\newcommand{\ese}{\end{subequations}}

\newcommand{\bv}{\bar{{\bf v}}}
\newcommand{\bu}{{\bf u}}
\newcommand{\rG}{{\rm G}}
\newcommand{\rM}{{\rm M}}
\newcommand{\rtf}{R}
\newcommand{\rnought}{R}
\newcommand{\energy}{\varepsilon}
\newcommand{\beps}{{\bm{\epsilon}}}
\input{epsf}

\begin{document}
\title{Vortices in Spatially Inhomogeneous Superfluids}
\author{Daniel E.~Sheehy and Leo Radzihovsky}
\affiliation{
Department of Physics, 
University of Colorado, 
Boulder, CO, 80309}
\date{June 8, 2004}
\begin{abstract}
  
  We study vortices in a radially inhomogeneous superfluid, as
  realized by a trapped degenerate Bose gas in a uniaxially symmetric
  potential.  We show that, in contrast to a homogeneous superfluid,
  an off-axis vortex corresponds to an anisotropic superflow
  whose profile strongly depends on the distance to the trap axis.
  One consequence of this superflow anisotropy is vortex precession
  about the trap axis in the absence of an imposed rotation.  In the
  complementary regime of a finite prescribed rotation, we compute the
  minimum-energy vortex density, showing that in the rapid-rotation limit
  it is extremely uniform, despite a strongly inhomogeneous
  (nearly) Thomas-Fermi condensate density $\rho_s(r)$.  The weak
  radially-dependent contribution ($\propto \nabla^2\ln\rho_s(r)$) to the vortex distribution, that
  vanishes with the number of vortices $N_v$ as 
  $\frac{1}{N_v}$,
  arises from the interplay between vortex quantum discretness (namely
  their inability to faithfully support the imposed rigid-body
  rotation) and the inhomogeneous superfluid density. This leads to
  an enhancement of the vortex density at the center of a typical
  concave trap, a prediction that is in quantitative agreement with
  recent experiments. One striking consequence of the inhomogeneous
  vortex distribution is an azimuthally-directed, radially-shearing
  superflow.

\end{abstract}
\maketitle


\section{Introduction}
\label{SEC:Intro}
An essential defining property of a superfluid is the way it responds
to an imposed rotation. Because a uniform superfluid state can only
support irrotational flow, one might expect a vanishing response to an
applied rotation in a simply-connected domain. However, dating back to
seminal works of Onsager and Feynman~\cite{onsager,feynman} it has
been understood that a superfluid can indeed rotate, but does so by
nucleating localized topological defects (vortices) that carry
discrete units of vorticity, thereby revealing the superfluid's
macroscopic quantum nature. In a finite system and for a sufficiently
low applied rotation rate $\Omega < \Omega_{c1}$, the energetic cost
of exciting even a single vortex is too high and the superfluid
remains stationary. For higher rates, vortices are nucleated into a
regular hexagonal lattice that, on average, approximates a uniform
response to the applied rigid-body rotation.

In recent years, rapid progress in the field of confined degenerate
Bose gases has led to the experimental realization of vortex lattices
containing large numbers of
vortices~\cite{Matthews,Madison,Haljan,Aboshaeer,Engels,Dalibard,JILA}.  Perhaps the
most striking feature of these lattices is their apparent {\em
  uniformity} despite the strong spatial variation of the local
superfluid density $\rho_s(r)$ imposed by the trap.  Despite some
recent attempts\cite{Anglin}, until very recently\cite{PRL} no
simple physical explanation for this uniformity (which has also been
observed in simulations~\cite{Feder01}) has appeared in the
literature.

To emphasize the apparent puzzle that such vortex uniformity presents,
one need not look further than the vortex state of a type-II
superconductor. There, vortices are well-known to be strongly pinned by
(i.e.~attracted to)  regions of suppressed superfluid density
created by material imperfections~\cite{commentSCII}. Based on this
analogy, one might expect that vortices in a confined Bose gas would
be repelled from the center of the trap (where superfluid density and
therefore vortex kinetic-energy cost is largest), and would congregate
near the condensate edges, resulting in a highly nonuniform and concave
vortex distribution.

In this Paper we present a theory of vortices in a confined
spatially-inhomogeneous rotating superfluid, providing a simple
physical explanation for and computing corrections to a uniform vortex
array. We explicitly calculate the vortex spatial distribution
$\bar{n}_v(r)$ in a trapped degenerate Bose gas characterized by 
$\rho_s(r)$, showing that it is given
by
\be
\label{fineomintro}
\bar{n}_v(r)\simeq \frac{m \Omega}{\pi \hbar} + \overline{c}\ \nabla^2
\ln \rho_s(r) , \ee
with $\overline{c}\equiv(8\pi)^{-1}\ln({\rm e}^{-1}/\xi ^2 \omega)$, $m$ the
boson mass, $\xi$ the vortex core size
and $\omega \equiv m\Omega/\hbar$ the scaled rotation
velocity. Since typically the spatial variation of the Thomas-Fermi (TF)
condensate density (or $\rho_s(r)$, see Ref.~\onlinecite{commentRhos}) takes place on the
scale of the condensate radius $R(\Omega)$, the $r$-dependent correction
in $\bar{n}_v(r)$ (second term) to the uniform rigid-body result
$n_{v0}\equiv m\Omega/\pi\hbar$ is subdominant in the thermodynamic and large
rotation limits, vanishing as $1/N_v$ with increasing number of
vortices.  More explicitly, for an isotropic harmonic trap, in the
Thomas-Fermi approximation we find
\be
\label{eq:tfintro}
\bar{n}_v(r) \simeq n_{v0} - \frac{1}{2\pi}
\frac{\rtf^2}{(\rtf^2 - r^2 )^2 } \ln \frac{{\rm e}^{-1}}{\xi^2 \omega}, \ee
where the condensate radius $R(\Omega)=R_0/\sqrt{1-(\Omega/\Omega_t)^2}$
is swelled by the centrifugal force beyond the Thomas-Fermi trap
radius $R_0$, diverging as the rotational velocity approaches the trap
frequency $\Omega_t$.

Hence we show that, indeed, consistent with
experiments\cite{Madison,Aboshaeer,Haljan,Engels,Dalibard,JILA},  the
vortex density is highly uniform and is well-approximated by the
rigid-body result. This uniformity can be most transparently
understood by ignoring vortex discreteness (valid at high rotation
rates\cite{commentQHE,BaymPethick}) and thinking in terms of the energetically
optimum superfluid velocity ${\bf v}_s(\br)$.  In a nutshell, for an
arbitrary smoothly-varying superfluid density $\rho_s(r)$, the
superfluid velocity of the rigid-body form ${\bf v}_s^0\equiv\Omega \zh\times \br$,
which corresponds to the uniform vortex density $n_{v0}=\frac{m}{h}\nabla\times
{\bf v}_s^0$, always minimizes the boson's London free-energy.  In
terms of vortices, this nearly uniform vortex distribution
$\bar{n}_v\approx n_{v0}$ 
is a consequence of a balance between the
spatial variation of the kinetic energy per vortex and the vortex
chemical potential, both of which scale with $\rho_s(r)$. 
While it is energetically costlier to position
vortices in a region where $\rho_s(r)$ is high (i.e.~the center of the
trap), the vortex chemical potential (controlled by $\rho_s(r)\Omega$)
is also high there, compensating and leading to an approximately
uniform vortex density.  The breakdown of the analogy with vortices in
type-II superconductors is therefore due to the difference in the
spatial dependence of the vortex chemical potential in the two cases.
While in superconductors this role is played by a {\em uniform}
external magnetic field $H$, in trapped superfluids the
vortex chemical potential is proportional to $\rho_s(r)$ and thus
spatially nonuniform.

\begin{figure}[bth]
\vspace{1.4cm}
\centering
\setlength{\unitlength}{1mm}
\begin{picture}(40,40)(0,0)
\put(-50,0){\begin{picture}(0,0)(0,0)
\includegraphics{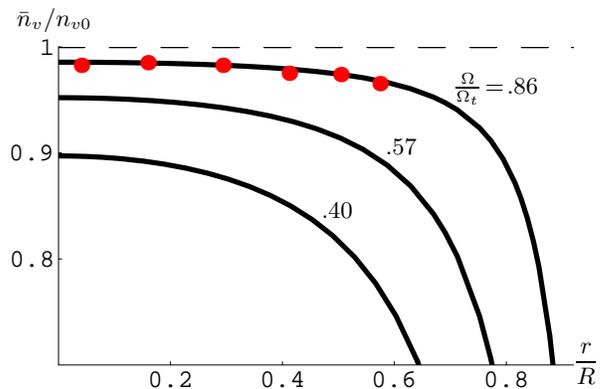}
\end{picture}}
\put(-17,50) {$\bar{n}_v/n_{v0}$}
\put(57,4.5) {$\displaystyle \frac{r}{R}$}
\put(41,41) {$\frac{\Omega}{\Omega_t}\! =\! .86$}
\put(32,33) {$.57$}
\put(23.5,24.5) {$.40$}
\end{picture}
\vspace{-.5cm}
\caption{(Color online) Plot of vortex density for the case of a 
  harmonic trap as a function of radius, Eq.~(\ref{eq:tfintro})
  (normalized to $n_{v0}$, dashed line), for the following
  $\Omega$ and $R$ values (labelled by the former): $\Omega = .86
  \Omega_t$ and $\rtf = 49 \mu m$; $\Omega = .57 \Omega_t$ and $\rtf =
  31 \mu m$; $\Omega = .40 \Omega_t$ and $\rtf = 25 \mu m$.  
  Points denote data adapted from I.  Coddington et al~\cite{JILA}.  }
\label{fig:TFplot}
\end{figure}

As we show here, a spatially-dependent correction to $\bar{n}_v(r)$ in
Eq.~(\ref{fineomintro}) arises from vortex discreteness and the related
inability of the vortex state to locally reproduce the uniform vorticity ${\bf v}_s^0$
corresponding to rigid-body rotation.
As shown rigorously long ago by Tkachenko~\cite{Tkachenko}, in the case of a {\em
  uniform} superfluid the resulting energetic frustration cannot be
reduced and the energetically optimum vortex distribution is a
regular vortex lattice with density $n_{v0}$.  In contrast, in an {\em inhomogeneous}
condensate the associated kinetic energy-density cost is
radially-dependent and can be lowered by a nonuniform vortex
distribution.
As illustrated in Fig.~\ref{fig:TFplot},
our analytical prediction for $\bar{n}_v(r)$, Eq.~(\ref{eq:tfintro}), is indeed
experimentally observable and shows remarkable agreement with recent
JILA experiments\cite{JILA}.  Furthermore, as we show below, our
prediction for the relation between $\bar{n}_v(r)$ and $\rho_s(r)$,
Eq.~(\ref{fineomintro}), can be more directly experimentally tested by
introducing a known inhomogeneity to $\rho_s(r)$ (by modifying the
trap potential) and studying the induced changes in $\bar{n}_v(r)$. Despite
the deviation of the vortex density from the rigid-body value, within
the London regime, the feedback effect on the condensate density $\rho_s(r)$ is
negligible, which remains well-approximated by the Thomas-Fermi
expression\cite{Watanabe,Cooper}.

\begin{figure}[bth]
\vspace{1.4cm}
\centering
\setlength{\unitlength}{1mm}
\begin{picture}(40,40)(0,0)
\put(-8,0){\begin{picture}(0,0)(0,0)
\includegraphics{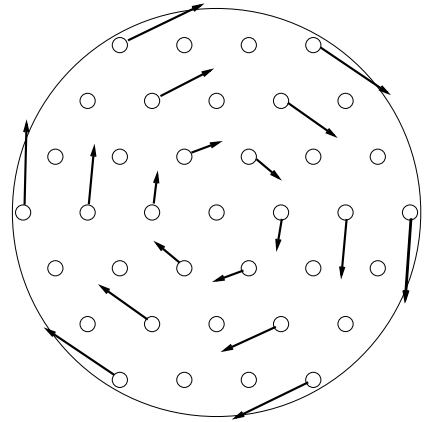}
\end{picture}}
\end{picture}
\vspace{-.3cm}
\caption{Schematic plot of the mean superfluid velocity (arrows) along 
with vortex positions (circles)
  in the frame rotating counter-clockwise at $\Omega$ (in which the {\it
    lattice is static\/}).  Thus, because ${\bf v}_s \neq \Omega \zh \times
  \br$ (and has a magnitude less than it), the superfluid flows past
  the lattice in the clockwise direction. }
\label{fig:vplot}
\end{figure}

One immediate consequence of our prediction of the deviation of
$\bar{n}_v(r)$ from the rigid-body value $n_{v0}$ is a finite superfluid
flow in the reference frame of the lattice (rotating with frequency
$\Omega$ relative to the lab frame). Even more interestingly, the
inhomogeneous vortex distribution $\bar{n}_v(r)$ implies that the resulting
azimuthal superflow in fact exhibits a radial shear, as schematically
illustrated in Fig.~\ref{fig:vplot}.

A prerequisite to the study of the thermodynamic vortex state under a
finite imposed rotation (which was our primary initial goal) is the
understanding of the single vortex problem. Although a number of
interesting results for the few-vortex problem have appeared in the
literature~\cite{Pismen,Fedichev,Linn,Lundh,McGee,Svidzinsky,Anglin02}, to
our knowledge no explicit, full solution for the velocity
${\bf v}_s(\br)=(\hbar/m)\nabla\theta(\br)$ around a vortex in an inhomogeneous
superfluid has been published.  Studying this problem in the London
approximation (valid deep in the superfluid state), we
find~\cite{Pismen,Svidzinsky,Svidzinskynote}, that the phase gradient at position $\br$ near a
vortex located off-axis at position $\br_0$ is given by
\bse
\bea
\label{eq:gradthetaintro} 
\grad \theta(\br) &=&  \frac{\zh \times (\br- \br_0)}{(\br - \br_0)^2}
+ \grad \theta_a(\br),
\\
\grad \theta_a(\br) &\simeq&
\frac{\zh \times \grad \rho_s(\br_0)}{2\rho_s(\br_0)}
\ln \frac{|\br -\br_0|}{R},
\label{eq:gradthetaintro2} 
\eea
\ese
with $\grad \theta_a(\br)$ giving the anisotropic correction to the
standard uniaxially-symmetric (homogeneous superfluid) result (first
term in Eq.~(\ref{eq:gradthetaintro})).  This additional {\em curl-free}
contribution to the superflow has interesting consequences for the
dynamics of individual vortices in an inhomogeneous superfluid.
Recall first that, in the absence of external forces, a vortex moves
with the local superfluid velocity.  For the case of a vortex in a
spatially inhomogeneous superfluid, $\grad \theta_a(\br)$ makes an
additional contribution to this superflow, leading to the remarkable
result (also discussed for specific trap potentials in
Refs.~\onlinecite{Fedichev,Linn,Lundh,McGee}) that a single
vortex at radius $r$ in an inhomogeneous superfluid will precess about
the trap's axis of symmetry with frequency
\be \omega_{\rm p} \approx \frac{\hbar}{m
  r}\frac{|\partial_r \rho_s|}{2\rho_s}\ln \frac{R}{\xi}. 
\ee
Such precession has been seen
experimentally in Ref.~\onlinecite{Anderson} with a quality factor of
order $10$.  We also demonstrate that a pair of vortices near the
center of the trap exhibit center-of-charge rotation about the trap
while also rotating about each other. 
The precession rate reassuringly vanishes for uniform $\rho_s$
in the thermodynamic limit, in which even an off-axis vortex is a
stationary state.

In the opposite limit, for $\br$ far away from the off-axis vortex at
$\br_0$, the additional contribution to the superflow induced by
condensate inhomogeneity has the form of a dipole field due to a
opposite-charge vortex at $\br_0$ and a vortex at the center of the
trap. Hence this analytical dipole contribution in the far field
effectively shifts the vortex to the center of the trap, leading to a
superflow that, far from the vortex, is axially symmetric, circulating
around the trap center and described by the phase-gradient
$\grad\theta=\frac{z\times \br}{r^2}$.

The rest of the paper is organized as follows. In Sec.~\ref{SEC:free}
we use a mean-field model (appropriate for our interests here) for a
rotating trapped Bose condensate to recall some standard results and
to derive the London equations for the superfluid velocity in the vortex
state.  In Sec.~\ref{SEC:analytic} we solve these equations to find the superflow
around a single off-axis vortex in an inhomogeneous condensate for a
number of experimentally-motivated condensate profiles.  We then use
these results in Sec.~\ref{SEC:ind} to derive precessional dynamics in one-
and two-vortex problems. In Sec.~\ref{SEC:lattice} we study the many-vortex
state and derive our primary result quoted in the Introduction, namely
the relation between the vortex and superfluid densities.  In
Sec.~\ref{SEC:elastic}, we derive a vortex lattice elastic energy in
an inhomogeneous condensate and use it to calculate the Tkachenko mode
equations (in the incompressible limit) for a spatially-varying
condensate profile. We conclude in Sec.~\ref{SEC:concluding} 
with a summary of our results and an outlook to future research.

\section{Model of a trapped (inhomogeneous) rotating superfluid}
\label{SEC:free}
The starting point of our analysis is the ground-state energy density of
an interacting rotating trapped Bose gas. To eliminate any explicit
time dependence, it is convenient to work in the frame of reference in
which the boundary conditions and the thermal cloud (the ``normal''
fluid, playing the role of the proverbial bucket) are
stationary~\cite{feynman}.  For simplicity, we focus on a trap with a
high degree of uniaxial anisotropy which reduces the problem to
two-dimensions perpendicular to the ($z$-) axis of rotation, although
our work can be straightforwardly generalized to three dimensions.
The corresponding boson energy density, written in a reference frame
rotating at frequency $\Omega$, is given by~\cite{mftCoherentZ}
\be \energy =\frac{\hbar^2}{2m} |\grad\Phi|^2 + (V({\bf r}) -\mu)
|\Phi|^2 + \frac{g}{2} |\Phi|^4 -\Omega L_z , \ee 
where the rotation frequency $\Omega$ plays the role of the ``chemical
potential'' for angular momentum $L_z$ with
\be
\Omega L_z = \Omega \Phi^{\dagger} [ -i\hbar \zh \cdot (\br \times
\grad)]\Phi. \ee
Here, $V(\br)$ is the trapping potential, $\mu$ is the boson number chemical
potential, and $g$ is  the s-wave scattering
potential.  Expressing the energy density in terms of the condensate
density $\rho_s(\br)$\cite{commentRhos} and phase $\theta(\br)$, defined
through the condensate field $\Phi(\br)=\sqrt{\rho_s}{\rm
  e}^{i\theta}$, (which microscopically is the
macroscopically-occupied single-particle wavefunction) we find
\bea
&&\energy =\frac{\hbar^2}{2m} [(\grad \sqrt{\rho_s})^2+ \rho_s(\grad \theta)^2 ]
+ (V(\br)-\mu) \rho_s + \frac{g}{2} \rho_s^2 
\nonumber \\
&&\quad+ i\hbar \Omega (\zh \times \br) \cdot
[\sqrt{\rho_s}\,\grad \sqrt{\rho_s} +i \rho_s \grad \theta].
\label{eq:f}
\eea
Deep within a dense Bose-condensed state, where $\rho_s(\br)$ is large,
it is convenient to eliminate the superfluid density from
Eq.~(\ref{eq:f}) by solving the corresponding Euler-Lagrange equation
\bea
\nonumber
&&0 = -\frac{\hbar^2}{2m} \big[ \nabla^2 \sqrt{\rho_s} -
\sqrt{\rho_s}(\grad \theta)^2 \big] + (V(\br) - \mu)\sqrt{\rho_s}
\\
&&\qquad 
 + g \rho_s^{3/2}- \hbar \Omega \sqrt{\rho_s} (\zh \times \br) \cdot \grad \theta .
\label{eq:rhoeom}
\eea
Neglecting derivatives in $\rho_s(r)$ and replacing $\grad \theta$ by
its approximate rigid-body value (see Sec.~\ref{SEC:rigid} below)
$\grad \theta \simeq \omega \zh \times \br$, we have the usual
Thomas-Fermi (TF) result~\cite{commentTF}
\be
\label{eq:tfrhopre}
\rho_s(\br) \approx (\mu - V(\br) +\frac{1}{2}m\Omega^2r^2)/g,
\ee
for the condensate density profile.  For the experimentally most
relevant case of a harmonic trap $V(r) = \frac{1}{2}m\Omega_t^2 r^2$
with $\Omega_t$ the trap frequency, Eq.~(\ref{eq:tfrhopre}) is 
uniaxially isotropic:
\be
\label{eq:tfrho}
\rho_s(r) \approx (\mu -\frac{1}{2}m(\Omega_t^2-\Omega^2)r^2)/g,
\ee
with the TF cloud radius $R(\Omega)$ (defined by the radius where
$\rho_s(r)$ vanishes) given by
\be
\label{eq:romega}
R(\Omega) = \frac{R_0}{\sqrt{1-(\Omega/\Omega_t)^2}},
\ee
and $R_0=\sqrt{2\mu/m\Omega_t^2}$ the cloud radius at zero rotation.
Using this solution of Eq.~(\ref{eq:rhoeom}) inside Eq.~(\ref{eq:f}) and
dropping unimportant constant terms we find the London
energy-functional for a rotating superfluid
\be
E =\frac{\hbar^2}{2m} \int d^2 r \rho_s(\br)[(\grad \theta)^2 
- 2\omega (\zh \times \br) \cdot\grad \theta],
\label{eq:f2}
\ee
with $\omega \equiv m\Omega/\hbar$.  

Under an imposed rotation, a superfluid turns by nucleating vortices.
In the London limit these are point singularities at a set of
positions $\br_i$ where the phase $\theta(\br)$ is nonanalytic and
satisfies
\bse
\bea
\label{eq:vortexsum}
&&\grad \times \grad \theta(\br) = 2\pi  n_v(\br)\zh,
\\
&&n_v(\br) = \sum_{i=1} \delta^{(2)} (\br-\br_i),
\label{eq:vortexsum2}
\eea
\ese
with $n_v(\br)$ the vortex density.
Eqs.~(\ref{eq:vortexsum}),(\ref{eq:vortexsum2}), together with the phase
Euler-Lagrange equation ($\frac{\delta E}{\delta \theta}=0$, with $E$ from
Eq.~(\ref{eq:f2})) 
\bse
\bea
\label{eq:fullthetaeqpre}
0&=& \grad\cdot(\rho_s(\br) \grad \theta) - 
\omega \grad \rho_s(\br)\cdot(\zh \times \br) ,
\\
\label{eq:fullthetaeq}
&=& \grad\cdot(\rho_s(\br) \grad \theta) ,
\eea
\ese
determine the superfluid phase $\theta(\br)$ and the corresponding
superfluid velocity ${\bf v}_s(\br)=\frac{\hbar}{m}\grad\theta$.
Here, Eq.~(\ref{eq:fullthetaeq}) applies for the case of a static trap
($\omega$ = 0) or, for a rotating trap in the experimentally relevant
case of uniaxial trap symmetry (i.e.~$\rho_s(\br) = \rho_s(r)$).
It is important to note that the vortex positions $\br_i(t)$ (and
therefore $n_v(\br)$) are static in the frame rotating with the vortex
lattice; thus, our time-independent expressions containing $n_v(\br)$
are defined in that frame.  We shall assume that vortices are static
in the frame rotating with the normal component, i.e., at frequency
$\Omega$.
%

\begin{figure}[bth]
\vspace{1.4cm}
\centering
\setlength{\unitlength}{1mm}
\begin{picture}(40,40)(0,0)
\put(-52,0){\begin{picture}(0,0)(0,0)
\includegraphics{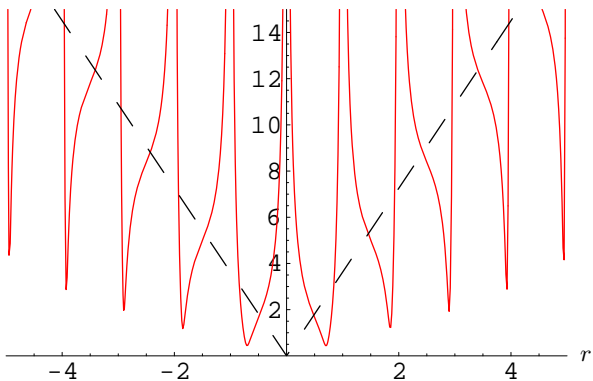}
\end{picture}}
\put(60,4.5) {$r$}
\end{picture}
\vspace{-.5cm}
\caption{(Color online) Magnitude of the phase gradient along a line intersecting vortices in
  a hexagonal rigid-body lattice (solid curve).  For comparison, the
  dashed curve shows the rigid-body result.  }
\label{fig:cores}
\end{figure}

\subsection{Rigid-body result}
\label{SEC:rigid}
As noted in the Introduction, the vortex discreteness embodied in
Eq.~(\ref{eq:vortexsum2}) makes the minimization of $E$, summarized by
Eqs.~(\ref{eq:f2}),(\ref{eq:vortexsum}) and (\ref{eq:fullthetaeq}),
 a nontrivial problem. 
For fast rotation (large $\Omega$, such that the coherence length $\xi$
is comparable to the vortex spacing $a$), the vortex state is dense and we can
neglect vortex discreteness, approximating ${\bf v}_s(\br)$ and
$n_v(\br)$ by arbitrary smooth functions.  Expressing $E$ in
Eq.~(\ref{eq:f2}) in terms of ${\bf v} _s({\bf r})$ (and dropping a
constant), we have
\be
\label{eq:f3}
E \simeq\frac{m}{2} \int d^2 r \rho_s(r)({\bf v}_s 
- \Omega (\zh \times \br))^2,
\ee
which is clearly minimized by the rigid-body solution 
\be
\label{eq:rigid}
{\bf v}_s = \Omega \zh \times \br, \ee 
corresponding to a {\it uniform\/} vortex density $n_{v0} =
\omega/\pi= m \Omega/ \pi \hbar$.

Clearly, this rigid-body result for ${\bf v}_s$ cannot be correct
for an arbitrary rotation rate $\Omega$ for the simple reason that it
corresponds to a uniform curl of $\grad \theta(\br)$, in contrast to
Eq.~(\ref{eq:vortexsum}).  This discrepancy can also be seen by
comparing a numerically computed magnitude of the exact superfluid
velocity for a uniform lattice of vortices to that for the rigid-body
result, Eq.~(\ref{eq:rigid}).  In Fig.~\ref{fig:cores} we plot $|\grad
\theta(\br)|$ satisfying Eq.~(\ref{eq:vortexsum}) for a hexagonal
array of vortices along a cut through several vortices, and compare it
to the rigid-body phase gradient plotted as a dashed line.

\subsection{Lower critical velocity of a rotating superfluid}
\label{SEC:omc1}
\begin{figure}
%
 \epsfxsize=\columnwidth
%
\centerline{\epsfbox{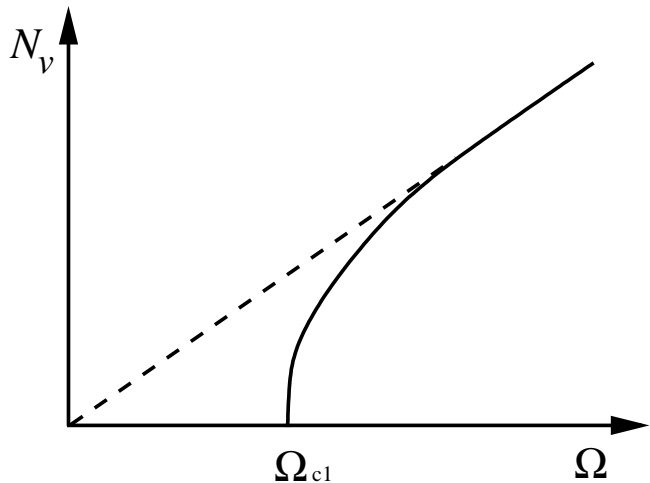}}
\caption{Schematic depiction of the number of vortices $N_v$ as a function of the applied rotation rate $\Omega$
  of a finite superfluid. The solid line is the expected actual number
  of vortices starting at the lower critical angular frequency
  $\Omega_{c1}$ whereas the dashed line is the rigid-body result (see
  main text).  }
\label{fig0}
\end{figure}

Before embarking on a proper calculation of the vortex density,
$n_v(\br)$, that correctly incorporates vortex discreteness, we recall
an elementary manifestation of the discrete nature of vortices in the
opposite limit of low $\Omega$, namely the existence of a
lower-critical (imposed) rotation rate $\Omega_{c1}$ below which the
superfluid ``refuses'' to rotate~\cite{leggett}.  To compute
$\Omega_{c1}$, we simply compare the energy Eq.~(\ref{eq:f2})
of a single vortex at the origin to the system's energy in the absence
of vortices.  The frequency where they cross is $\Omega_{c1}$. For
$\Omega>0$, the solution of Eqs.~(\ref{eq:vortexsum}),(\ref{eq:fullthetaeq})
 for a vortex at the origin leads to
a counter-clockwise circulation:
\be
\grad \theta = \frac{\hat{\phi}}{r}.
\label{eq:single}
\ee
For simplicity, we take for $\rho_s(\br)$ a simple form that is given by
a constant $\rho_0$ inside a radius $\rnought$ and zero outside this
radius.  Thus, inserting Eq.~(\ref{eq:single}) into Eq.~(\ref{eq:f2})
and evaluating the integral we have
\be
E = \frac{\hbar^2 \rho_0}{2m}  [ 2\pi \ln \frac{\rnought}{\xi} - 2 \pi \omega \rnought^2
],
\label{omc1}
\ee 
where $\xi$ is the size of the vortex core derivable from
Eq.~(\ref{eq:rhoeom}) in the presence of a vortex
 and providing a short-distance cutoff for London
theory.  Since clearly for the case of no vortex $E=0$, we find a
well-known result for a superfluid confined to radius $R$ (see, e.g.,
Ref.~\onlinecite{Baym96})
\be
\label{eq:omc}
\Omega_{c1} = \frac{\hbar}{m\rnought^2} \ln\frac{\rnought}{\xi},
\ee
which, unlike the analogous problem of type-II superconductors
(because of the absence of screening, i.e., an infinite London
penetration length), vanishes in the thermodynamic limit.  Since
vortices do not appear for $\Omega < \Omega_{c1}$, Eq.~(\ref{eq:omc})
gives the first hint of the violation of the rigid-body result. This
suggests that even for $\Omega > \Omega_{c1}$ the actual number of
vortices $N_v(\Omega)$ is expected to be {\it less\/} than the rigid-body result to
which it must asymptotically approach
 in the large $\Omega$ limit.
Therefore, based on these general arguments we expect $N_v(\Omega)$ to follow the solid curve in
Fig.~\ref{fig0}.\cite{commentDiscreteN,Butts} 

Thus far, we have provided elementary calculations exhibiting the
limiting behavior of a rotating confined Bose gas in the extreme
large\cite{commentQHE} and small $\Omega$ regimes.  Before turning to
the general many-vortex problem, in the next section, we first study the
(prerequisite) single vortex problem in a spatially inhomogeneous
superfluid.

\section{Single vortex in a spatially inhomogeneous superfluid}
\label{SEC:analytic}

In spite of its importance to understanding vortex states in
inhomogeneous rotating condensates, the single vortex problem in an
trapped rotating BEC has received relatively little attention, with
most treatments using the standard homogeneous-condensate vortex
solution as a starting
point and focusing on 
dynamics~\cite{Pismen,Fedichev,Linn,Lundh,McGee,Svidzinsky,Anglin02}. Two
notable exceptions are works by Rubinstein and Pismen~\cite{Pismen}
and by Svidzinsky and Fetter\cite{Svidzinsky}.
However, because of the focus in Ref.~\onlinecite{Svidzinsky} on the
single-vortex dynamics, only asymptotics of the superfluid velocity near
an off-axis vortex was worked out. On the other hand, although an
exact solution for a single vortex was found in
Ref.~\onlinecite{Pismen}, it corresponds to a very special (i.e.~not
motivated by any experimental realization) $\rho_s(\br)$.

In contrast, our interest in computing the vortex density distribution
(Sec.~\ref{SEC:lattice}) requires a full analysis of the superflow corresponding to
a single off-axis vortex in an inhomogeneous condensate and this is
the subject of this section.  To determine the correct superfluid
velocity, we study the Euler-Lagrange equation for $\theta$
Eq.~(\ref{eq:fullthetaeq}) (which expresses the conservation of the
boson number current) along with the vorticity constraint for a vortex
at $\br_0$:
\be 
\label{eq:constraint}
\grad \times \grad \theta = 2 \pi \delta^{(2)} (\br - \br_0). 
\ee
For the most general problem, these equations must be supplemented by
a boundary condition $\hat{n}\cdot(\rho_s(\br)\grad \theta)|_R=0$ enforcing
the vanishing of boson current through the boundary of the system.
Consequently, the energy and dynamics of a vortex are strongly
affected by both the condensate inhomogeneity and the nontrivial boundary
conditions~\cite{commentBucket}. However, 
for the case of a Bose gase trapped by a smooth
potential (relevant to experiments), the condensate density
$\rho_s(\br)$ vanishes at the cloud's boundary, automatically
satisfying the above vanishing current condition.

Before analyzing Eqs.~(\ref{eq:fullthetaeq}) and (\ref{eq:constraint})
in detail, we note that (inspired by the vortex solution
$\grad\theta=\frac{\zh \times (\br -\br_0)}{(\br -\br_0)^2}$ for a
uniform superfluid), an exact solution to Eq.~(\ref{eq:fullthetaeq})
is given by
\be
\label{eq:ansatz}
\grad \theta(\br) 
= \frac{\rho_s(\br_0)}{\rho_s(\br)}\frac{\zh \times (\br -\br_0)}{(\br -\br_0)^2}.
\ee
Although this solution fails to satisfy Eq.~(\ref{eq:constraint}) 
(satisfying it only to leading order in spatial gradients of
$\rho_s(\br)$) and therefore is not a proper vortex solution, it {\it
  does\/} exhibit the qualitatively correct behavior that we shall
verify in this section: $\grad \theta$ (and therefore the superfluid
velocity) is generally larger where $\rho_s(\br)$ is small and smaller
where $\rho_s(\br)$ is large, relative to the solution for a uniform
$\rho_s$.

For a full and systematic analysis, it is convenient to represent
\begin{equation}
\label{eq:theta}
\theta(\br) = \theta_v(\br) + \theta_a(\br),
\end{equation}
in terms of the known singular part
\bse
\bea
\label{eq:thetav}
\theta_v(\br)&=& \tan^{-1}\left[\frac{y-y_0}{x-x_0}\right],\\
&=&\phi ,
\label{eq:thetav2}
\eea 
\ese 
corresponding to a vortex for a uniform superfluid
density, with
\begin{equation}
\grad \theta_v(\br)= \frac{\zh\times(\br-\br_0)}{(\br-\br_0)^2},
\label{eq:gradthetav}
\end{equation}
that ensures the vortex carries the correct topological charge in
Eq.~(\ref{eq:constraint}).  The analytic part $\theta_a(\br)$ ($\grad
\times \grad \theta_a(\br)=0$) is chosen so that $\theta(\br)$
satisfies the Euler-Lagrange equation Eq.~(\ref{eq:fullthetaeq}),
which reduces to
\begin{equation}
\label{eq:elanalytic}
\grad\rho_s \cdot  \grad \theta_a +  \rho_s \nabla^2 \theta_a  
= - \grad \rho_s\cdot \grad \theta_v.
\end{equation}
(Henceforth in this section we are restricting attention to 
the uniaxial symmetry case $\rho_s(\br) = \rho_s(r)$.)
A virtue of this approach is that it reduces the problem to
the  analytical
solution of Eq.~(\ref{eq:elanalytic}) subject to a known \lq\lq
source\rq\rq\ field $-\grad\rho_s\cdot\grad \theta_v(\br)$,
Eq.~(\ref{eq:gradthetav}).  In the case of a uniform $\rho_s$,
Eq.~(\ref{eq:elanalytic}) is obviously solved by $\grad \theta_a(\br)=
0$, with $\grad \theta = \grad \theta_v(\br)$ reducing to the
well-known result Eq.~(\ref{eq:gradthetav}).
 In the presence of a nonzero gradient of
$\rho_s(r)$, $\grad \theta_a(\br)$ generally provides a nontrivial
(but curl-free) correction to the superfluid velocity in
Eq.~(\ref{eq:gradthetav}).  In addition, since for a smoothly-varying
$\rho_s(r)$ we expect that $\grad \theta$ is {\it approximately\/}
given by the uniform result Eq.~(\ref{eq:gradthetav}), this
formulation is naturally set up for a systematic expansion in the small
parameter $\grad\ln\rho_s$.

Finally we note that for a vortex at the center of the trap ($\br_0 =
0$), $\grad \rho_s \propto - \rh$  leads to a vanishing of the
source term $-\grad\rho_s\cdot\grad \theta_v(\br)=0$. Consequently for
an on-axis vortex $\grad \theta_a(\br)= 0$ and the solution reduces
to a simple axially-symmetric result $\grad \theta =\grad
\theta_v=\frac{\zh\times\br}{r^2}=\grad\phi$.

Thus, below we naturally focus on the nontrivial case of an
off-trap-axis ($\br_0\neq0$) vortex. Although for a generic
$\rho_s(r)$ no closed form solution to Eq.~(\ref{eq:elanalytic}) is
available, a systematic asymptotic analysis in all relevant regimes
determined by three length scales (the condensate size $R$, the vortex
position $\br_0$ and the displacement from the vortex $\br - \br_0$)
is possible and is presented in Secs.~\ref{SEC:gauss} and
~\ref{SEC:smooth} (near a vortex $|\br - \br_0| \ll r_0\ll R$) and in
Sec.~\ref{SEC:outer} (away from the vortex $r_0\ll|\br - \br_0|$).
\begin{figure}[bth]
\vspace{1.4cm}
\centering
\setlength{\unitlength}{1mm}
\begin{picture}(40,40)(0,0)
\put(-50,0){\begin{picture}(0,0)(0,0)
\includegraphics{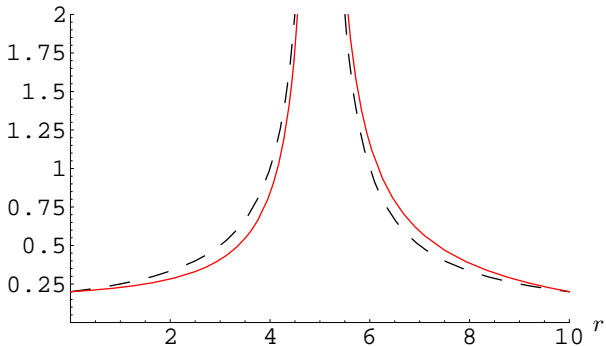}
\end{picture}}
\put(59,4.5) {$r$}
\end{picture}
\vspace{-.5cm}
\caption{(Color online)   Phase-gradient magnitude (solid curve)
  through a vortex away from the center of the trap, for the case of
  the Gaussian condensate profile Eq.~(\ref{eq:gaussian}).  The vortex
  is located at spatial position $\br_0 = (5,0)$ with $\rnought = 5$
  characterizing the scale of the condensate.  For comparison, the
  dashed curve depicts $|\grad \theta_v|$ only, neglecting the analytic field.
}
\label{fig:gausscross2}
\end{figure}
\begin{figure}[bth]
\vspace{1.4cm}
\centering
\setlength{\unitlength}{1mm}
\begin{picture}(40,40)(0,0)
\put(-50,0){\begin{picture}(0,0)(0,0)
\includegraphics{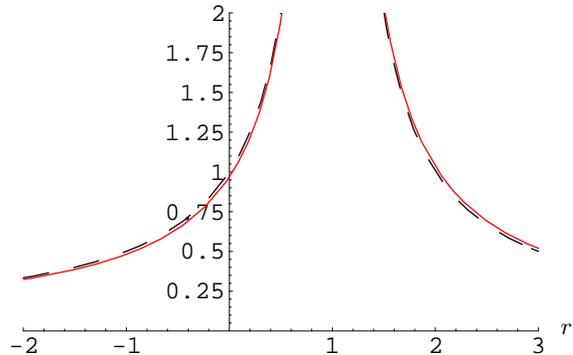}
\end{picture}}
\put(57,4.5) {$r$}
\end{picture}
\vspace{-.5cm}
\caption{(Color online) Phase-gradient magnitude (solid curve) through  
  a vortex near the center of the trap, for the case of the Gaussian
  condensate profile Eq.~(\ref{eq:gaussian}).  The vortex is located
  at spatial position $\br_0 = (1,0)$ with $\rnought = 5$
  characterizing the scale of the condensate.  For comparison, the
  dashed curve depicts $|\grad \theta_v|$ only, neglecting the analytic field.
}
\label{fig:gausscross1}
\end{figure}
\subsection{Gaussian condensate profile}
\label{SEC:gauss}

Before studying the general case of an arbitrary condensate profile
$\rho_s(r)$, here we first analyze Eq.~(\ref{eq:elanalytic}) for the
simpler special case of a Gaussian condensate profile
\be
\label{eq:gaussian}
\rho_s(r) = \rho_0\exp(-r^2/2\rnought^2),
\ee
with $\rnought$ characterizing the size of the condensate.  For this
choice, Eq.~(\ref{eq:elanalytic}) reduces to
\be
-\br\cdot   \grad \theta_a + \rnought^2\nabla^2 \theta_a  = \br \cdot \grad \theta_v.
\label{eq:response2p}
\ee

Focusing on the most interesting regime, we solve
Eq.~(\ref{eq:response2p}) in the vicinity of the vortex.  Close to the
vortex at $\br_0$, we can neglect the first term on the left side
of Eq.~(\ref{eq:response2p}) in comparison to the second, $-\br \cdot
\grad \theta_a + \rnought^2\nabla^2 \theta_a \simeq \rnought^2
\nabla^2\theta_a$, reducing the equation for $\theta_a$ to
\be
\nabla^2 \theta_a = \frac{1}{\rnought^2}\grad \theta_v \cdot \br .
\label{eq:response2}
\ee
A solution to Eq.~(\ref{eq:response2}) is then easily found to be
\be
\label{eq:thetaa2fin}
\theta_a(\br) =\frac{1}{2\rnought^2}(\br-\br_0)\cdot(\br_0 \times \zh
) \ln \frac{|\br-\br_0|}{\rnought}, 
\ee 
where the denominator of the argument of the logarithm amounts to a
choice of integration constant, fixed by matching to an ``outer''
solution.  Our approximate choice above may be motivated physically by
noting that we expect $\theta_a$ to exhibit spatial variation on a
scale given by $\rnought$, as will be confirmed by more general
calculations to follow.  Although $\br_0 \cdot(\br_0 \times \zh ) =
0$, we have written Eq.~(\ref{eq:thetaa2fin}) in this way to emphasize
that close to the vortex, the $\br$ dependence in $\theta_a$ arises
through the $\br - \br_0$ combination only.

To emphasize key qualitative features of the solution $\theta(\br)$,
we reexpress the vortex solution in terms of the azimuthal angle $\phi$
between $\br - \br_0$ and $\br_0$. In terms of $\phi$ the singular
part of the solution is (recall Eq.~(\ref{eq:thetav2})) simply $\theta_v = \phi$.  From
Eq.~(\ref{eq:thetaa2fin}) the analytic contribution $\theta_a(\br)$
can be seen to be given by $\theta_a(\phi) = c \sin \phi$, leading to
\begin{equation}
\theta(\phi) = \phi + c \sin \phi,
\end{equation}
with $c=\frac{1}{2R^2}r_0|\br - \br_0|\ln \frac{R}{|\br-\br_0|}$.

An important (required) feature of this solution is that $\theta_a(\phi)$
is indeed analytic, carrying vanishing winding $\int_0^{2\pi} d\phi
\,\partial_\phi \theta_a(\phi) = 0$, and therefore preserving the quantization in
units of $2\pi$ of the winding of the full vortex phase
$\theta(\phi)$.  The consequence of the nontrivial analytic
contribution is that $\theta(\phi)$ obeys this quantization condition
by winding more slowly near the center of the trap (i.e.~where $\phi
\approx \pi$) than away from it (i.e.~where $\phi \approx 0$) in
agreement with our intuitive approximate vortex solution
Eq.~(\ref{eq:ansatz}).

As we illustrate in Fig.~\ref{fig:gausscross2}, the analytical
contribution $\theta_a$ has  a small but qualitatively important effect
on the superfluid velocity in the vicinity of a vortex. We compare the
vortex superfluid velocity magnitude $|\grad \theta|$ for a vortex in
an inhomogeneous condensate to its counterpart $|\grad \theta_v|$ in
the homogeneous condensate along $\xh$ through the vortex located
at $\br_0=x_0\xh$. While both velocity profiles exhibit the
characteristic $1/r$ divergence, consistent with our general arguments
the superfluid velocity $|\grad \theta|$ is clearly smaller than
$|\grad \theta_v|$ near the center of the trap and larger than it away from
the center of the trap, with this effect vanishing (see
Fig.~\ref{fig:gausscross1}) for a vortex near the center of the trap.

\subsection{Generic spatially-varying condensate profile: ``inner'' solution}
\label{SEC:smooth}
We now turn to the full problem of finding the superfluid velocity near
a vortex in an inhomogeneous superfluid defined by a general
(uniaxially symmetric and smooth, with $R\gg \xi$) 
$\rho_s(r)$.

Our method is a simple generalization of the calculation for the
Gaussian case in Sec.~\ref{SEC:gauss} and applies in the regime $|\br
- \br_0| \ll r_0$.  Working to lowest order in the gradient of the
condensate profile, near a particular vortex we approximate the smoothly
varying $\rho_s(r)$ and its gradient $\grad\rho_s(r)$ by their values
at the vortex position $\br_0$. This considerably simplifies the
Euler-Lagrange equation (while retaining the essential physics) to
\be
\label{eq:unapprox1}
 \grad \mu(r_0) \cdot  \grad \theta_a(\br) +  
\frac{1}{2}\nabla^2 \theta_a(\br) = -  \grad \mu(r_0) \cdot \grad \theta_v,
\ee
where we defined
\be
\label{eq:mudef}
  \grad \mu(r ) \equiv\frac{1}{2}\frac{\grad\rho_s(r)}{\rho_s(r)}.
\ee
The solution to Eq.~(\ref{eq:unapprox1}) can now be easily obtained by
first finding a corresponding Green function that satisfies
\be \big( \grad \mu(r_0) \cdot \grad + \frac{1}{2}\nabla^2\big)
G(\br-\br') = \delta^{(2)}(\br-\br').
\label{eq:green}
\ee
An explicit expression for $G(\br)$ (verifiable via direct
substitution) may be expressed in terms of the Bessel function
$K_0(x)$:
\be
G(\br) = -\frac{1}{\pi} {\rm e}^{-\br \cdot \grad\mu(r_0)} K_0(r|\grad\mu(r_0)|).
\ee
This then leads to the solution to Eq.~(\ref{eq:unapprox1}) given by
\bse
\bea
\label{eq:thetasol1}
\hspace{-2cm}\theta_a(\br) &=& \! \!-\int d^2 r' G(\br -\br') \grad\mu(r_0)
 \cdot \frac{\zh\times(\br'-\br_0)}{(\br'-\br_0)^2},
\\
&=&\! \!-  \grad\mu(r_0)\! \cdot \!\int d^2 r' G(-\br')  
\frac{\zh\times(\br'+\br-\br_0)}{(\br'+\br-\br_0)^2},
\label{eq:thetasol2}
\eea
\ese
where in going to Eq.~(\ref{eq:thetasol2}) we have shifted the
integration variable $\br' \to \br' + \br$, with the resulting
integral on the right side clearly a function of $\br - \br_0$ only (a
consequence of our approximation above).  Moreover, since the second
factor is sharply peaked near $\br' \simeq \br_0 - \br$, it is valid
to this order of approximation to replace $G(-\br')$ by its value at
this point.  The remaining integral may be easily evaluated, finally
yielding
\bse 
\bea 
\hspace{-2cm}\theta_a(\br)&\approx& - (\br - \br_0) \cdot ( \zh
\times \grad \mu(r_0) ) {\rm e}^{-(\br -\br_0 ) \cdot \grad
  \mu(r_0) }
\nonumber \\
\hspace{-2cm}&&\times K_0(|\br - \br_0||\grad \mu(r_0|) ,
\label{eq:firstapprox} 
\\
&\approx& (\br - \br_0) \!\cdot\! ( \zh \times \grad \mu(r_0) ) 
\ln |\br - \br_0| |\grad \mu(r_0)|,
\label{eq:firstapprox2} 
\eea
\ese
where in Eq.~(\ref{eq:firstapprox2}) we have taken the $\br \to \br_0$
limit of Eq.~(\ref{eq:firstapprox}).  Since $\mu(r)$ varies on the
scale of the condensate size $R$, the analytical correction $\theta_a$
scales like $1/R^2$, and, as expected, vanishes in the thermodynamic
limit.

Using $\grad \mu(r_0) = -\rh_0 |\partial_r\mu(r_0)|$ we find
that at fixed distance near the vortex, the azimuthal $\phi$
dependence of $\theta(\phi)$ is given by
\begin{equation}
\theta(\phi)\approx \phi+c \sin\phi\ {\rm e}^{c'\cos\phi},
\label{phase}
\end{equation}
with $c=|\partial_r\mu(r_0)||\br - \br_0|K_0(|\br -\br_0||\partial_r\mu(r_0)|)$ and 
$c'=|\br - \br_0||\partial_r\mu(r_0)|$ positive $\phi$-independent functions
that increase with increasing $|\br - \br_0|$ (in the stated small 
$|\br - \br_0|$ regime).  Because of the smallness of $c'$
near $\br_0$, for a Gaussian condensate profile this general result
reduces to that found in Sec~\ref{SEC:gauss}. Again, as required by
analyticity of $\theta_a(\br)$, it is easy to show that $\int_0^{2\pi} d\phi
\,\partial_\phi \theta_a(\phi) = 0$, thereby preserving quantization of circulation
of superflow encoded in $\int_0^{2\pi} d\phi \,\partial_\phi \theta(\phi) = 2\pi$.

Although the phase (and corresponding superflow)
distortion given by $\theta(\phi)$ is quite small (especially near the vortex) we note that
there are a number of experimental techniques that have been used
succesfully to measure the condensate phase growth around a
vortex~\cite{Matthews,Inouye}, giving hope that the prediction in
Eq.~(\ref{phase}) may be experimentally testable.

\begin{figure}[bth]
\vspace{.4cm}
\centering
\setlength{\unitlength}{1mm}
\begin{picture}(40,40)(0,0)
\put(-20,-10){\begin{picture}(0,0)(0,0)
\includegraphics{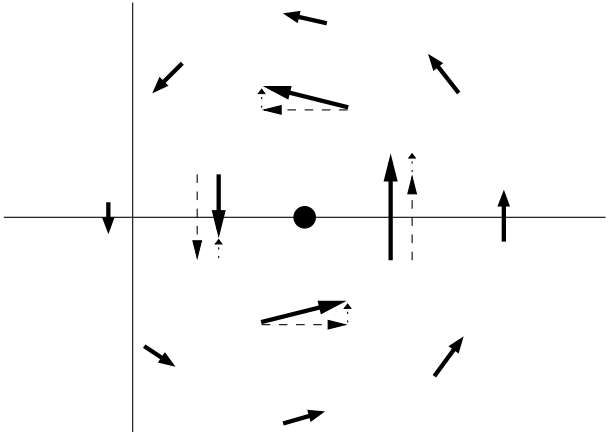}
\end{picture}}
\end{picture}
\vspace{.7cm}
\caption{Schematic picture of the effect of the analytic field on the superfluid velocity of
an off-axis vortex (Eq.~(\ref{eq:approxgrad})); the vortex location is denoted by a filled circle.  
The solid arrows denote $\grad \theta$ as 
affected by the nearly uniform analytic field.  On the innermost ring the  nearly uniform 
$\grad \theta_a$ is depicted as a small dotted arrow and $\grad \theta_v$ is depicted as
a  larger dashed arrow; their vector sum leads to the distorted $\grad \theta$ that is 
smaller in magnitude near the center of the trap and larger near the edge.}
\label{fig:cartoon}
\end{figure}

Using Eq.~(\ref{eq:firstapprox2}), the superfluid velocity near  a vortex (defined by $|\br- \br_0|
|\grad \mu(r_0)| \ll 1$) is easily computed, giving
\bea
&&{\bf  v}_s(\br)\approx\frac{\hbar}{m}
\Big[\frac{\zh\times(\br-\br_0)}{(\br-\br_0)^2}
\nonumber \\
&&\qquad \qquad +
\zh \times\grad \mu(r_0) \ln(|\br - \br_0| |\grad \mu(r_0)|)\Big],  \label{eq:approxgrad}
\eea
in agreement with the result found in Ref.~\onlinecite{Pismen}.  Hence
quite generally, near a vortex the correction to the superfluid
velocity arising from the inhomogeneity of the condensate is
approximately spatially uniform and is orthogonal to the displacement
$\br_0$ from the trap center.  We see that, as anticipated based on general
arguments above, the superflow is slower at $\phi=\pi$ and faster at
$\phi=0$ (see Fig.~\ref{fig:cartoon}).  Moreover, for $\br_0 \to 0$,
$\grad \mu(r_0)\to 0$, so that the analytic field is negligible for a vortex 
near the center of the trap.
%

%
\subsection{Generic spatially-varying condensate profile: ``outer'' solution}
\label{SEC:outer}

We now calculate the asymptotics of the superfluid velocity far away
from the vortex.  Interestingly, after some examination of the form of
Eq.~(\ref{eq:elanalytic}) it is clear that the outer solution can be
easily found. To see this first note that for $\theta_a$ that
satisfies $\nabla^2\theta_a = 0$ Eq.~(\ref{eq:gradthetav}) reduces to
\begin{equation}
\grad\rho_s\cdot\grad\theta_a = -\grad\rho_s\cdot\grad\theta_v.
\end{equation}
Naively this would be satisfied by the antivortex (at $\br_0$) solution
\bse
\bea
\grad\theta_a&=&-\grad\theta_v,\\
&=&-\frac{\zh\times(\br-\br_0)}{(\br-\br_0)^2},
\label{naiveouter}
\eea \ese 
(which satisfies the assumed condition $\nabla^2\theta_a = 0$) were it
not for the analyticity requirement $\oint d\br \cdot \grad \theta_a(\br) = 0$.  
However, this is easily fixed by adding to the 
solution in Eq.~(\ref{naiveouter}) another contribution
$\frac{\zh\times\br}{r^2}$ due to a positive vortex located at the
origin, which clearly satisfies the homogeneous part of
Eq.~(\ref{eq:elanalytic}). We thus obtain an {\em exact} outer
solution\cite{commentFixBC} (verifiable by direct substitution)
to Eq.~(\ref{eq:elanalytic}):
\be
\label{eq:large}
\grad \theta_a(\br) = \frac{\zh \times \br}{r^2} -  \frac{\zh \times (\br- \br_0)}{(\br - \br_0)^2}.
\ee
Because the solution Eq.~(\ref{eq:large}) corresponds to a vortex dipole,
with a positive vortex at the trap center and a negative vortex at the
location of the true vortex $\br_0$ it is easy to see that, as
required, it satisfies the vanishing circulation
($\grad\times\grad\theta_a=0$) condition for $|\br - \br_0| \gg r_0$.
Hence far from an off-axis vortex at $\br_0$, the superfluid velocity
is given by
\begin{equation}
{\bf v}_s(\br)\simeq \frac{\hbar}{m}\frac{\zh \times \br}{r^2},
\end{equation}
i.e., the superflow adjusts to be axially symmetric about the trap
center, such that a vortex at $\br = \br_0$ {\it appears} to be
sitting at $\br = 0$.
%
\begin{figure}[bth]
\vspace{1.4cm}
\centering
\setlength{\unitlength}{1mm}
\begin{picture}(40,40)(0,0)
\put(-45,0){\begin{picture}(0,0)(0,0)
\includegraphics{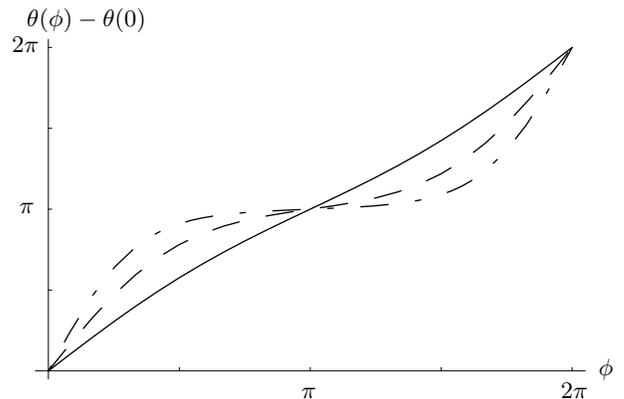}
\end{picture}}
\put(-17,48) {$ \theta(\phi) - \theta(0)$}
\put(59,1.5) {$\phi$}
\put(54,-1.5) {$2\pi$}
\put(19.5,-1.5) {$\pi$}
\put(-19,44.5) {$2\pi$}
\put(-18,23) {$\pi$}
\end{picture}
\vspace{.2cm}
\caption{Plot of the phase around a vortex in an inhomogeneous
  superfluid for the Bessel profile Eq.~(\ref{eq:bessel}) as a
  function of the angle $\phi$ between $\br - \br_0$ and $\br_0$, for
  Eq.~(\ref{eq:pismen}) integrated numerically.  The parameters $R =
  10$ and $r_0 = 20$ and the curves represent $|\br -\br_0| = 1$
  (solid), $10$ (dashed) and $100$ (dot-dashed).  }
\label{fig:phase1}
\end{figure}
\begin{figure}[bth]
\vspace{1.4cm}
\centering
\setlength{\unitlength}{1mm}
\begin{picture}(40,40)(0,0)
\put(-45,0){\begin{picture}(0,0)(0,0)
\includegraphics{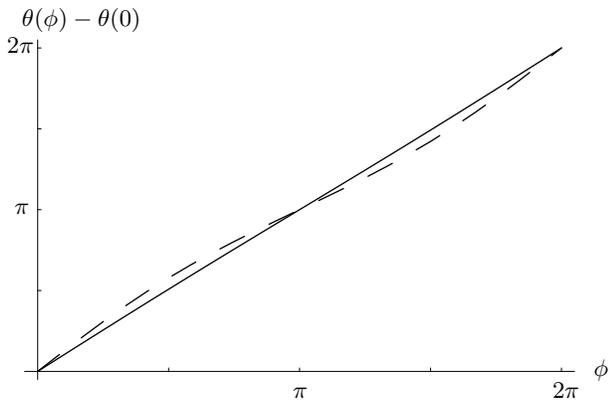}
\end{picture}}
\put(-17,48) {$ \theta(\phi) - \theta(0)$}
\put(59,1.5) {$\phi$}
\put(53.5,-1.5) {$2\pi$}
\put(19.,-1.5) {$\pi$}
\put(-19,44.5) {$2\pi$}
\put(-18,23) {$\pi$}
\end{picture}
\vspace{.2cm}
\caption{Same as Fig.~\ref{fig:phase1} but with $\rnought = 20$, $r_0 = 5$ and $|\br - \br_0| = 1$ (solid line) and
  $|\br - \br_0| = 1000$ (dashed line) exhibiting the smallness of the analytic
  field for vortices close to the center of the trap.  }
\label{fig:phase2}
\end{figure}
\begin{figure}
%
 \epsfxsize=\columnwidth
%
\centerline{\epsfbox{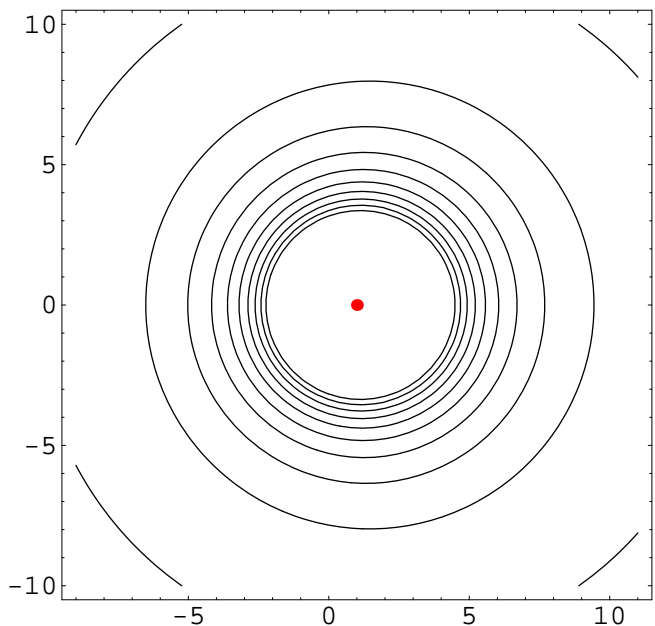}}
\caption{(Color online)
  Contour plot of constant phase gradient curves for a vortex at
  $\br_0 = (1,0)$ with $\rnought = 10$. Their near-circular symmetry
  indicates the unimportance of the analytic field in this regime.  }
\label{fig:contour1}
\end{figure}
\begin{figure}
%
  \epsfxsize=\columnwidth
%
\centerline{\epsfbox{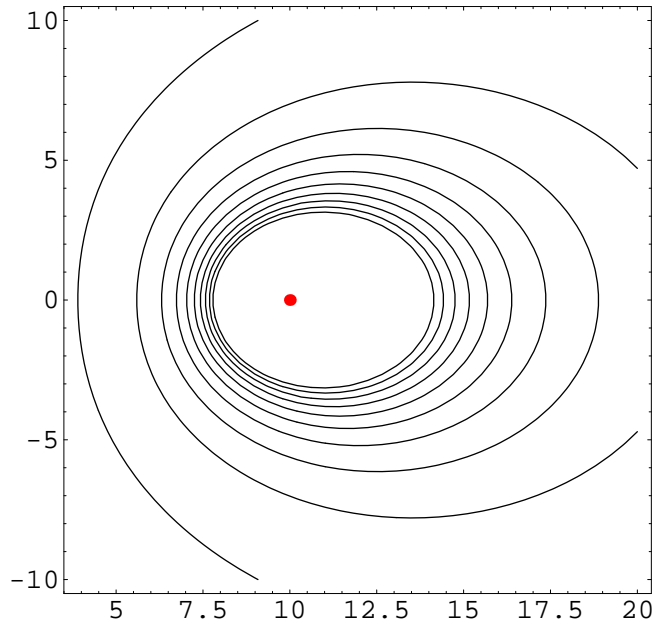}}
\caption{(Color online)
  Contour plot of constant phase gradient curves for a vortex at
  $\br_0 = (10,0)$ with $\rnought = 10$.  The effect of the analytic
  field is clearly seen in the distorted shapes of these curves; the
  phase gradient has been decreased to the left of the vortex and
  increased to the right.  }
\label{fig:contour2}
\end{figure}

\begin{figure}[bth]
\vspace{1.4cm}
\centering
\setlength{\unitlength}{1mm}
\begin{picture}(40,40)(0,0)
\put(-50,0){\begin{picture}(0,0)(0,0)
\includegraphics{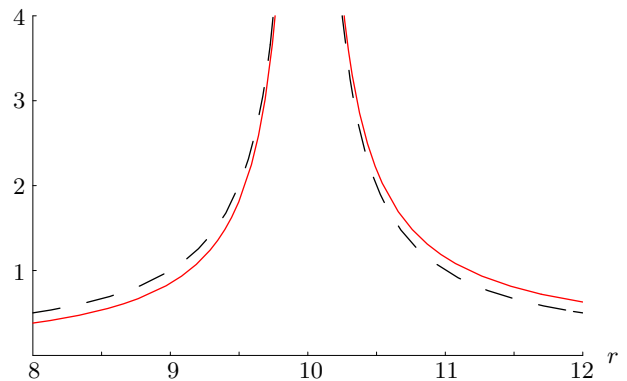}
\end{picture}}
\put(60,0) {$r$}
\put(55,-2) {$12$}
\put(37,-2) {$11$}
\put(19,-2) {$10$}
\put(1.5,-2) {$9$}
\put(-17,-2) {$8$}
\put(-19,45) {$4$}
\put(-19,34) {$3$}
\put(-19,22.5) {$2$}
\put(-19,11) {$1$}
\end{picture}
\caption{(Color online) 
  Phase-gradient magnitude (Eq.~(\ref{eq:pismen})) through a vortex
  far from the center of the trap, located at spatial position $\br_0
  = (10,0)$ with $\rnought = 10$.  For comparison, the dashed curve
  depicts $|\grad \theta_v|$ only, neglecting the analytic field
  (solid curve).  }
\label{fig:pismencross}
\end{figure}

\subsection{Bessel condensate profile: Exact solution due to Rubinstein and
  Pismen}
\label{SEC:Pismen}
Although we have been unable to find a closed form vortex solution for
an arbitrary condensate profile $\rho_s(r)$, for a specially contrived
but qualitatively realistic profile
\be
\label{eq:bessel}
\rho_s(r) = \frac{\rho_0}{I_0(r/\rnought)^2}, 
\ee
a closed form solution was found by Rubinstein and Pismen in the
context of optical vortices in an inhomogeneous laser
beam~\cite{Pismen}.  In Eq.~(\ref{eq:bessel}), $I_0(x)$ is the modified
Bessel function, with the asymptotic behavior $I_0(x) \to 1+x^2/4 $
and $\exp(x)/\sqrt{2\pi x}$ for $x\to 0$ and $x\to \infty$,
respectively.  This asymptotics shows that this special $\rho_s(r)$
behaves as $1 - r^2/2\rnought^2$ for $r\ll \rnought$ (i.e.~it is
parabolic) and decays exponentially for $r\gg \rnought$.  Thus,
despite its fine-tuned very specific form, it is qualitatively similar
to the experimentally relevant TF profile.

Although most of the properties of a vortex are contained in the
asymptotic solutions found in previous sections, it is useful to check
the predictions found there for a general $\rho_s(r)$ against the
specific but exact solution for $\rho_s(r)$ in Eq.~(\ref{eq:bessel}).
This is the subject of present subsection. 

Following Ref.~\onlinecite{Pismen}, the Euler-Lagrange equation
(\ref{eq:fullthetaeq}) can be solved exactly by introducing an
auxilary function $\chi$ related to $\theta$ by $ \rho_s(r) \grad
\theta=\zh\times\grad\chi$.  With this transformation and the very
special choice of $\rho_s(r)$ in Eq.~(\ref{eq:bessel})
the vortex quantization condition Eq.~(\ref{eq:constraint}) reduces
to a simple Helmholtz-like partial differential
equation for $\rho_s^{-1/2}\chi(r)$, which in turn
leads to the solution~\cite{Pismen}
\bea \nonumber &&\grad \theta(\br) = -{\rm e}^{\mu(r_0) - \mu(r)} \zh
\times \big[K_0(|\br -\br_0|/\rnought) \grad \mu(r)
\\
&& \qquad \qquad + \grad K_0(|\br -\br_0|/\rnought)
\big],\label{eq:pismen} \eea
where $\mu(r) = - \ln I_0(r/\rnought)$ and $K_0(x)$ is the Bessel
function. In the regime $r,r_0 \ll R$ of primary interest to us we
find (using $K_0(x) \approx -\ln x$, for $x\to 0$)
\bea
\nonumber
&&\grad \theta(\br) \simeq   \frac{\zh \times \rh}{R} 
 \frac{I_1(r/\rnought)}{I_0(r_0/\rnought)}\ln \frac{R}{|\br -\br_0|}
\\
&& \qquad \qquad + \frac{\zh \times (\br - \br_0)}{(\br -\br_0)^2}
  \frac{I_0(r/\rnought)}{I_0(r_0/\rnought)}.\label{eq:pismen2}
\eea
For $\br \to \br_0$, the second term dominates inside
$\grad\theta$, leaving the first term as a small correction.  In this
regime, the first curl-free term is the $\grad \theta_a$ analytic
contribution introduced in Eq.~(\ref{eq:theta}).  Thus, this exact result
agrees with the general one found in Eq.~(\ref{eq:approxgrad}) and
discussed in the Introduction, Eq.~(\ref{eq:gradthetaintro}).

In the opposite regime, far away from the vortex ($r \gg r_0$), but
still away from the condensate boundary ($r \ll R$), the first term
continues to be subdominant to the ``bare'' vortex contribution
(second term). However, its $\br$ dependence can no longer be ignored
and leads to a superflow that is perpendicular to the position vector
$\br$, rather than the vortex position vector $\br_0$. 

Even further in the far field, defined by $r\gg \rnought$ but $r_0\ll
\rnought$ (so that the vortex is still near the center of the
trap)\cite{comment_farR}, the exact solution Eq.~(\ref{eq:pismen})
reduces simply to
\be \grad\theta(\br) \approx \frac{\zh \times \br}{r^2} , \ee
with small power-law corrections in $R/r$, in agreement with our \lq\lq
outer\rq\rq\ solution Eq.~(\ref{eq:large}).

The vortex distortion due to the condensate inhomogeneity can also be seen in
the angular coordinate depedence of $\theta(\phi)-\theta(0)$ (obtained by
numerically integrating Eq.~(\ref{eq:pismen}) along a contour between
$0$ and $\phi$) that we plot in Figs.~\ref{fig:phase1} and~\ref{fig:phase2} for different
radii $|\br - \br_0|$.  As we found in Sec.~\ref{SEC:gauss}, the
distinction between the exact vortex $\theta(\phi)$ and
$\theta_v(\phi)=\phi$ is largest for vortices far from the center of
the trap (see Fig.~\ref{fig:phase1}), and grows with $|\br - \br_0|$. 
For a vortex near the center of the trap (Fig.~\ref{fig:phase2}), the distinction is barely noticeable.
From examining these
 plots it is clear that an
off-axis vortex exhibits an azimuthal dependence of the phase that (as
found in Sec.~\ref{SEC:smooth}) is given by $\theta(\phi) \approx \phi
+ c \sin\phi$ with $c$ a positive constant that grows with distance
away from the vortex, $|\br - \br_0|$, and with distance of the vortex
from the trap axis, $r_0$.

To further depict the increasing importance of the analytic field for
vortices located away from the center of the trap, in
Figs.~\ref{fig:contour1} and \ref{fig:contour2} we plot (for $\rnought = 10$) 
contours of
constant $|\grad \theta|$ for a vortex at $\br_0 = (1,0)$ and $\br_0 =
(10,0)$, respectively.  As expected from earlier
discussion, Eq.~(\ref{eq:approxgrad}), the superflow of a vortex near the center of the
trap (Fig.~\ref{fig:contour1}) characterized by nearly circular
contours contrasts strongly with that of a far-off-axis vortex
(Fig.~\ref{fig:contour2}) described by highly distorted contours of
contant superfluid velocity.  This latter distortion is a manifestation
of our earlier finding of a larger correction to the
superfluid velocity for vortices away from the
center of the trap and smaller near it. In Fig.~\ref{fig:pismencross} we depict 
the phase-gradient magnitude (solid curve) for the same parameters as
Fig.~\ref{fig:contour2}, contrasting it with that of a vortex in a
homogeneous condensate (dashed curve) given by $|\grad \theta_v|$.

\section{Precessional dynamics of individual vortices in spatially inhomogeneous superfluids}
\label{SEC:ind}
Having found in Sec.~\ref{SEC:analytic} the  vortex superflow profile
in a spatially inhomogeneous superfluid, here we study implications of
these results for few-vortex dynamics. In the absence of rotation
($\Omega= 0< \Omega_{c1}$), vortex excitations are {\it metastable\/},
and therefore should eventually escape the condensate, allowing the
superfluid to relax to the ground state.  However, because of the
(massless) guiding-center dynamics of the vortex, which conserves angular
momentum, we expect the vortex lifetime to be quite long (limited by
the trap's azimuthal asymmetry and decoherence). As we show in the next
subsection, on shorter timescales a vortex in a confined Bose gas,
located away from the trap axis, {\it necessarily\/} precesses around the
center of the trap\cite{Linn,Lundh,McGee}.  This can be understood from
general considerations by noting that energy eigenstates in an
axially-symmetric (finite) trap are also angular momentum eigenstates.
Consequently, an off-axis point vortex (that clearly is not an
eigenstate of angular momentum) is not an energy eigenstate and must
therefore evolve at constant energy. As we demonstrate in 
Appendix~\ref{SEC:appendix} such vortex precession is a property of any confined
superfluid, even with a uniform condensate density.

\subsection{Precession of a single vortex}

We shall now demonstrate (via two different routes) that an off-axis
vortex in a spatially inhomogeneous superfluid precesses around the
trap center\cite{Linn,Lundh,McGee}.  To begin, we recall that the dynamics of
a vortex is governed by the Magnus
``force''~\cite{Ambegaokar,Ao}
\be {\bf F}_{\rM} = 2\pi \hbar \rho_s \zh\times (\dot{\br}_i -
{\bf v}_s) ,
\label{eq:magnus0}
\ee
where $\dot{\br}_i$ is the velocity of the $i$th vortex and
${\bf v}_s$ is the background superfluid velocity (not
including the symmetric divergent part $\grad \theta_v$ due to the vortex itself) at the
location of the vortex, $\br_i$.  In the absence of other forces and
in equilibrium ${\bf F}_{\rM} = 0$, giving that a vortex moves at the
local superfluid velocity, $\dot{\br}_i = {\bf v}_s$.  In an
inhomogeneous superfluid, it is appropriate to identify 
${\bf v}_s$ with the curl-free contribution $\hbar \grad \theta_a/m$
that, as discusssed in Sec.~\ref{SEC:analytic}, arises from condensate
inhomogeneity.  Thus, using the second (analytic) term of 
Eq.~(\ref{eq:approxgrad}) with $|\br
-\br_i| \approx \xi$, we find
\be
\label{eq:singlevortdyn}
\dot{\br}_i \simeq \frac{\hbar}{m} \zh \times \grad \mu(r_i) \ln \xi |\grad\mu(r_i)|,
\ee
where we remind the reader that $\grad \mu(r)$ is given by Eq.~(\ref{eq:mudef}).

The result in Eq.~(\ref{eq:singlevortdyn}) may be also obtained by
computing the energy $E(\br_i)$ of a vortex (corresponding to the kinetic
energy of the superfluid) at position $\br_i$ in an inhomogeneous
superfluid.  Balancing the associated force against the Magnus
``force''~\cite{McGee} gives the equation of motion of the vortex:
\be
2\pi \hbar \rho_s \zh\times \dot{\br}_i - \grad E(\br_i) = 0,
\label{eq:magnusgrad}
\ee
where the curl-free part of the vortex flow $\grad \theta_a$ is now 
regarded as part of the vortex and therefore the external superflow
${\bf v}_s$ in Eq.~(\ref{eq:magnusgrad}) is taken to be zero.
Starting with the kinetic energy, Eq.~(\ref{eq:f2}) (at $\omega = 0$),
expressing it in terms of $\theta_v$ and $\theta_a$, and using the
equation of motion for $\theta_a$, Eq.~(\ref{eq:elanalytic}), we find
\be
\label{eq:energyind}
E(\br_i) =\frac{\hbar^2}{2m} \int d^2 r [\rho_s(r)(\grad \theta_v)^2 -
\rho_s(r)(\grad \theta_a)^2], 
\ee
with $\theta_a$ the solution to Eq.~(\ref{eq:elanalytic}) and $\grad\theta_v$
given by Eq.~(\ref{eq:gradthetav}). Because the curl-free superfluid
velocity correction $\frac{\hbar}{m}\grad \theta_a$ arises from the
condensate inhomogeneity, its contribution to the vortex kinetic
energy is subdominant, vanishing as $1/R^2$. In the dominant (first)
term the superfluid velocity $\frac{\hbar}{m}\grad \theta_v$ diverges
at $\br_i$, allowing us to ignore the $\br$ dependence of the smoothly
varying density $\rho_s(r)$ and approximate it by its value at the
vortex, $\rho_s(r)\simeq \rho_s(r_i)$. This leads to
\be
\label{eq:energyind2}
E(\br_i) \approx \frac{\hbar^2}{2m} \rho_s(r_i) 2 \pi \ln
\frac{R}{\xi}, 
\ee 
which is correct to leading order in the small parameter $\xi/R$.
When inserted into Eq.~(\ref{eq:magnusgrad}), we have
\be
\label{eq:singlevortdyn2}
\dot{\br}_i \simeq -\frac{\hbar}{m} \zh \times \grad \mu(r_i) \ln \frac{R}{\xi},
\ee
in agreement (up to weak logarithmic corrections) with
Eq.~(\ref{eq:singlevortdyn}) found above.  

For an axially symmetric trap, with $\grad \mu = \rh \partial_r
\mu(r)$, the equation of motion Eq.~(\ref{eq:singlevortdyn2}) can be easily
solved and (as advertised) gives vortex precession about the center of
the trap:
\bea
\label{eq:indrot} 
\br_i(t) &=& r_i (\xh \cos \omega_{\rm p} t + \yh \sin
\omega_{\rm p} t),
\eea
with the direction of rotation in the same sense as vortex circulation
(i.e., counterclockwise here) and with the angular frequency $\omega_{\rm
  p}$ given by:
\bse
\bea
\omega_{\rm p} &=& \frac{\hbar}{m r_i} \partial_r \mu(r_i)
\ln\frac{\xi}{R},
\label{eq:indrotfreq} 
\\
&=&\frac{\hbar}{m r_i} \frac{\partial_r \rho_s(r_i)}{2\rho_s(r_i)} \ln
\frac{\xi}{R}.
\label{eq:indrotfreq2} 
\eea
\ese
For simplicity, in Eq.~(\ref{eq:indrot}) we have chosen a particular
initial condition $\br_i(0) = \xh r_i$.  For the experimentally
relevant case of a TF condensate profile $\rho_s(r) = \rho_0
(1-r^2/R^2)$, the precession frequency Eq.~(\ref{eq:indrotfreq2})
reduces to
\be
\omega_{\rm p} =  \frac{\hbar}{m } \frac{1}{R^2 - r_i^2} \ln \frac{R}{\xi},
\label{eq:indrotfreq3} 
\ee
which agrees qualitatively with JILA experiments that observe such
precession at angular frequency $\approx \hbar/m R^2$ that is roughly
independent of $r_i$~\cite{Anderson}.

\subsection{Precession of multiple vortices}

The above analysis can be easily extended to the dynamics of many vortices,
where each vortex moves with the local superfluid velocity that is due
to the sum of its own curl-free superflow $\frac{\hbar}{m}\grad\theta_a$ (generated by the condensate 
inhomogeneity) and
the 
superflow of all other vortices.  For the simplest case of two
vortices the equations of motion are given by
\bse
\bea
\label{eq:paireom1}
\dot{\br}_1  &=&  \frac{\hbar}{m}\grad \theta_{a,1}(\br_1) + \frac{\hbar}{m} \grad\theta_2(\br_1;\br_2),
\\
\dot{\br}_2  &=&  \frac{\hbar}{m}\grad \theta_{a,2}(\br_2) + \frac{\hbar}{m} \grad\theta_1(\br_2;\br_1),
\label{eq:paireom2}
\eea \ese
where $\grad \theta_{a,i}(\br_i)$ is the analytic phase gradient
induced by vortex $i$ at position $\br_i$ and $\grad
\theta_i(\br_j;\br_i)$ is the {\it total\/} phase gradient induced at
position $\br_j$ due to vortex $i$ which is located at position
$\br_i$.  Equations~(\ref{eq:paireom1}) and (\ref{eq:paireom2}) are
quite general.
We now restrict attention to the case of a TF condensate profile and
focus on the limit $r_i \ll R$ so that the rotation frequency
Eq.~(\ref{eq:indrotfreq3}) is approximately independent of radius.  As
we showed in Sec.~\ref{SEC:analytic}, $\grad \theta_a$ is almost
always much smaller than $\grad \theta_v$ (especially when $r_i \ll
R$) so that the second terms in Eq.~(\ref{eq:paireom1}) and
Eq.~(\ref{eq:paireom2}) can be well-approximated by neglecting the
associated analytic fields.  Under these conditions we have
\bse
\bea
\label{eq:paireom1p}
\dot{\br}_1  &=&  \frac{\hbar}{m}\Big( \frac{1}{R^2} \zh \times \br_1 \ln\frac{R}{\xi} 
+ \frac{\zh \times (\br_1 - \br_2)}{(\br_1 -\br_2)^2}\Big),
\\
\dot{\br}_2  &=&
\frac{\hbar}{m}\Big( \frac{1}{R^2} \zh \times \br_2 \ln\frac{R}{\xi} 
+ \frac{\zh \times (\br_2 - \br_1)}{(\br_2 -\br_1)^2}\Big).
\label{eq:paireom2p}
\eea
\ese
The two equations decouple when we change variables to relative
($\brho \equiv \br_1 - \br_2$) and center-of-charge (${\bf R}_c \equiv
(\br_1 + \br_2)/2$) coordinates:
\bse
\bea
\label{eq:paireomrel}
\dot{\brho}  &=&  \frac{\hbar}{m}\Big(\frac{\ln R/\xi}{R^2} \zh \times 
\brho 
+ 2\frac{\zh \times \brho}{\rho^2}\Big),
\\
\dot{{\bf R}}_c  &=&
\frac{\hbar}{m R^2}\ln\frac{R}{\xi} \zh \times {\bf R}_c,
\label{eq:paireomcm}
\eea
\ese
so that the vortex pair center-of-charge rotates around the origin at
the lower-critical frequency $\Omega_{c1}$
\be
\omega_c = \frac{\hbar}{m R^2}  \ln\frac{R}{\xi},
\ee
and the two vortices orbit each other at frequency
\be
\omega_\rho = \frac{\hbar}{m}\Big(\frac{1}{R^2}  \ln\frac{R}{\xi} + \frac{2}{\rho^2}\Big),
\ee
which increases rapidly with decreasing vortex separation $\rho$, with the
divergence at $\rho\rightarrow0$ (as usual) cut off by the coherence
length $\xi$.

\section{Vortex array in a spatially inhomogeneous superfluid}
\label{SEC:lattice}

We now turn to the many-vortex problem, with the goal of computing the
vortex spatial distribution in an inhomogeneous rotating superfluid.
The superfluid velocity ${\bf v}_s=\frac{\hbar}{m}\grad\theta$, measured
in the laboratory frame, due to an array of $N$ vortices is a solution
of Eqs~(\ref{eq:vortexsum}) and (\ref{eq:vortexsum2})
 from Sec.~\ref{SEC:free}. By linearity of these equations it
is given as the sum of the contributions from each vortex:
\be
\grad\theta(\br)=
\sum_{i=1}^N \frac{\zh \times (\br -\br_i)}{(\br -\br_i)^2},
\label{eq:sum}
\ee
with vortex positions $\br_i$ static in the frame of the normal
component (rotating with frequency $\Omega$ relative to the lab
frame). In the above, we have justifiably neglected the curl-free
$\grad\theta_a$ contribution to the superflow of a vortex, studied in
Sec.~\ref{SEC:analytic}, that, as can be seen from 
Eq.~(\ref{eq:approxgrad}), scales as
$\grad\rho_s/\rho_s\sim 1/R$ and therefore gives a contribution that is
subdominant (in our expansion in $\grad\rho_s/\rho_s$) in the large
condensate limit. Moreover, since $\grad \theta_a$ is curl-free it cannot
contribute to the rotation of the superfluid and thus is expected to 
be irrelevant for the rapid-rotation limit.
The corresponding vortex density is given by Eq.~(\ref{eq:vortexsum2})
\be
n_v(\br)=(2\pi)^{-1}\grad\times\grad\theta=\sum_i^N\delta^{2}({\bf
  r}-\br_i).
\label{eq:sumn_v}
\ee

As discussed in Sec.~\ref{SEC:rigid}, for high rotation rates
$\Omega$ the vortex state is dense, and to a high accuracy we can
neglect the discrete nature of vortices [embodied by
Eqs.~(\ref{eq:sum}),(\ref{eq:sumn_v})] and approximate the superfluid
velocity ${\bf v}_s(\br)$ and vortex density $n_v(\br)$ by arbitrary
smooth functions that minimize the total energy, Eq.~(\ref{eq:f2}).  As
we showed in Sec.~\ref{SEC:rigid}, within this continuum approximation,
the superfluid velocity is simply given by the rigid-body result ${\bf
  v}_s=\Omega\zh\times\br$ and $n_{v0}=m\Omega/\pi\hbar$,
thereby providing a simple explanation for the high vortex lattice
uniformity observed in experiments\cite{Madison,Aboshaeer,Haljan,Engels,Dalibard,JILA}.  
Despite this
agreement, it is of interest to understand the degree of accuracy and
limitations of this uniform vortex distribution (rigid-body)
prediction.

Away from this classical rapid-rotation limit\cite{commentQHE}, vortex
discreteness begins to matter and the rigid-body uniform vortex
distribution solution clearly breaks down, as ${\bf v}_s(\br)$
diverges as $1/|{\bf r}-\br_j|$ near each vortex at $\br_j$. As
illustrated in Fig.~\ref{fig:cores}, ${\bf v}_s(\br)$ thus strongly deviates
from rigid-body flow. In this regime, where a superfluid exhibits
its locally irrotational quantum nature, the summation in
Eq.~(\ref{eq:sum}) can no longer be replaced by an integration, and
the minimization of $E$ must be done directly over vortex positions,
$\br_i$, rather than over a field ${\bf v}_s(\br)$. In fact, as we saw
in Sec.~\ref{SEC:omc1}, this vortex discreteness manifests itself even in a 
uniform but finite-size condensate through the lower-critical
rotational velocity $\Omega_{c1}\approx(\hbar/m R^2)\ln\frac{R}{\xi}$
below which no rotation is supported by the condensate.

For a {\em uniform} infinite condensate the problem was solved long
ago by Tkachenko~\cite{Tkachenko}, who rigorously demonstrated that
the solution is a hexagonal lattice characterized by the vortex
density $n_{v0}$. Carrying out Tkachenko's exact analysis for a
spatially-varying $\rho_s(r)$ is a formidable task.  We shall instead
develop an approximate continuum theory that nevertheless incorporates
the essential vortex discreteness and which is valid for a smooth condensate
profile $\rho_s(r)$, with accuracy controlled by $\grad\ln\rho_s$.

\subsection{Vortex lattice in a generic inhomogeneous superfluid
  density profile}
\label{SEC:phasegrad}

We have shown that to compute the vortex distribution in an inhomogeneous condensate,
it is essential to faithfully incorporate vortex {\it discreteness\/}
in treating the sum in Eq.~(\ref{eq:sum}).  For ${\bf r}$ near a
vortex located at $\br_j$, the superflow is clearly dominated (see
Fig.~\ref{fig:cores}) by a diverging contribution from the $j$th vortex, with
other vortices making a subleading and {\it smoothly varying\/}
correction to $\grad \theta(\br)$.  To formalize this, we write $\br =
\br_j + \delta \br$ and split the sum in Eq.~(\ref{eq:sum}) into a
contribution from the $j$th vortex plus a contribution due to all
remaining vortices:
\be
\label{eq:second}
\grad \theta (\br_j + \delta\br) =
\frac{\zh \times \delta\br}{\delta r^2} + 
\zh \times \sum_{i\neq j} \frac{\br_j + \delta \br -\br_i}{(\br_j + \delta\br -\br_i)^2}.
\ee
In the smooth correction to the superflow due to all other vortices
the sum over vortex positions can be safely approximated by an
integral over a continuously distributed vortex density:
\bse
\bea
\label{eq:tkachenko2}
\hspace{-1cm}\frac{\hbar}{m} \grad\theta(\br_j +\delta\br)&\simeq&
\frac{\hbar}{m}\frac{\zh \times \delta\br}{\delta r^2} + 
\bv_s(\br_j+\delta\br),
\\
\bv_s(\br) &\equiv &
\frac{\hbar}{m}\int d^2 r'\, \bar{n}_v(\br') \frac{\zh\times(\br -\br')}{(\br -\br')^2},
\label{eq:tkachenko2p}
\eea
\ese
where $\bar{n}_v(\br)$ and $\bv_s(\br)$ are the vortex density and
superfluid velocity at position $\br$, coarse-grained on the scale of the
lattice spacing.  
Equation~(\ref{eq:tkachenko2p}) may be simplified by expressing the
vortex coordinate $\br_i = \br_i^0 + \bu(\br_i^0)$ with the 
$\br_i^0$ forming a uniform hexagonal lattice at average
density $n_{v0}$ and $\bu(\br)$ 
being a static vortex displacement field.  In terms of $\bu(\br)$,
\be
\label{eq:nvu}
 \bar{n}_v(\br) \simeq n_{v0} (1-\grad \cdot \bu(\br)) ,
\ee
so that the integral in
Eq.~(\ref{eq:tkachenko2p}) gives
\be
\label{eq:ve}
\bv_s(\br) =\Omega [ \zh \times \br - 2\zh \times \bu_L(\br) ].  \ee
Thus, the coarse-grained part of the superflow is only sensitive to
the {\em longitudinal} vortex displacement, related to $\bu$ through
\bse
\bea
\bu_L(\br) &=& \frac{\grad\grad}{\nabla^2}\cdot\bu,\\
&=& \int d^2 r' \grad G(\br -\br') \grad \cdot \bu(\br'),
\label{eq:long}
\eea
\ese
with $G(\br -\br')$ the Green function for the Laplacian.  Since (as
we will see below) for the case of main interest of an
axially-symmetric trap the optimum vortex lattice distortion is
purely longitudinal, we may take $\bu_L = \bu$.
Equation~(\ref{eq:tkachenko2}) is a remarkable result that illustrates
how vortices, each with a singular $\hat{\phi}/r$ superfluid velocity,
add up to approximate rigid-body flow\cite{commentTkachenko} with
the (second) $u$-term characterizing deviations from it.

To compute the vortex distribution $\bar{n}_v(\br)$, we express the energy
Eq.~(\ref{eq:f2}) in terms of the vortex lattice displacement field
$\bu(\br)$ and minimize it over $\bu(\br)$.  To this end, we express the
energy $E$ of an array of vortices as a sum over contributions due to
individual unit cells, with each cell associated with a particular
vortex\cite{baym1,Fischer}:
\be
E= \frac{\hbar^2}{2m}\sum_{i} \!\int_{i} d^2 r \rho_s(\br) [\grad \theta(\br)
  - \omega (\zh\times \br )]^2,
\label{eq:cellfree}
\ee 
where the subscript $i$ on the integral indicates that it is to be
performed over a unit cell centered at $\br_i$.  
In obtaining Eq.~(\ref{eq:cellfree}), we have completed the square in 
Eq.~(\ref{eq:f2}) and discarded a constant term.
Using
Eq.~(\ref{eq:tkachenko2}) for the phase gradient within the $i$th cell,
shifting the integration in each  cell via $\br \to \br + \br_{i}$
and using the fact that $\rho_s(\br)$ does not vary appreciably over a
cell (i.e.~$\rho_s(\br + \br_{i}) \simeq \rho_s(\br_{i})$ inside a
particular cell), we find
\be
E \simeq  \frac{\hbar^2}{2m}\sum_{i}  \rho_s(\br_{i})
\int_{i} d^2 r  
\big[
\frac{\zh \times \br}{r^2} - 2\omega \zh \times \bu_L(\br +\br_j)
\big]^2.
\label{eq:preint}
\ee
The dominant contribution to the energy per cell comes from the
diverging superfluid velocity at the center of a cell; thus the way in
which we treat the cells' boundary is unimportant.  Taking each cell to be
a circle of radius $a$, set by the average vortex density
$\bar{n}_v(\br)=1/\pi a^2$ and approximating the smoothly-varying
field $\bu_L(\br)$ by its value at the vortex position $\bu_L(\br +\br_i)
\approx \bu_L(\br_i)$, the integrals over the cell areas 
can be easily done, giving
\be
E\simeq  \frac{\hbar^2}{2m}\sum_i \rho_s(\br_i ) \Big[ \pi \ln \frac{1}{\pi\xi^2 \bar{n}_v}
+\frac{4\omega^2 u_L (\br_i)^2}{\bar{n}_v(\br_i)}
\Big],
\label{eq:int}
\ee
with the short-scale logarithmic divergence of the vortex kinetic energy
as usual cut off by the vortex core size $\xi$.

The vortex discreteness effects that we have emphasized, arising from
the singular nature of the phase gradient near the core of a vortex,
are contained within the first term of the energy functional
Eq.~(\ref{eq:int}) and clearly vanish at high vortex density as $\pi
\xi^2 \bar{n}_v\rightarrow 1$.\cite{commentQHE} The remaining sum over
vortex cells can be faithfully approximated by an integral
\bse 
\bea
\sum_i\ldots &=& \int d^2 r \,n_v(\br)\ldots,\\
&\approx& \int d^2 r \bar{n}_v(\br)\ldots,
\label{sum_i}
\eea
\ese
which, after using Eq.~(\ref{eq:nvu}), finally gives the energy
functional of a vortex solid in an inhomogeneous rotating condensate:
\bea 
\nonumber 
&&\hspace{-.5cm}E[\bu(\br)]\simeq  \frac{\hbar^2}{2m}\int d^2 r \rho_s(\br)\Big[ 4\omega^2 u_L(\br)^2
\\
&&\qquad \quad+ \omega(1- \grad \cdot \bu(\br)) \ln \frac{1}{\xi^2 \omega (1-\grad \cdot \bu )}
\Big].\label{eq:finf} \eea

Armed with $E[\bu]$, the vortex distribution can be easily computed by
minimizing $E[\bu]$ over $\bu(\br)$, namely by solving $\frac{\delta
  E}{\delta \bu} = 0$.  Since Eq.~(\ref{eq:finf}) depends only on the
longitudinal component\cite{commentUL} $\bu_L$ (recall $\grad \cdot
\bu(\br) = \grad \cdot \bu_L(\br)$), we can equivalently vary with
respect to $\bu_L$.  Thus, we obtain the nontrivial ground-state
distortion of a vortex lattice from the naive uniform (rigid-body)
state in an inhomogeneous condensate characterized by the uniaxially symmetric
superfluid
density $\rho_s(\br) = \rho_s(r)$:
\be 
\bu(\br) = -\frac{1}{8 \omega \rho_s(r) } \grad \Big[ \rho_s(r) \ln
\frac{c}{\xi^2 \omega(1-\grad \cdot \bu)} \Big],
\label{eq:uresult}
\ee
where $c \equiv 1/{\rm e}$. For the case of $\rho_s(r)$ largest at the
center of the trap, the vortex distortion $\bu(\br)$ is clearly in the
outward radial direction and purely longitudinal for the case of
an axially symmetric $\rho_s(r)$. 

The nontrivial distortion $\bu(\br)$ arises as a result of the competition
between the vortex kinetic energy (second term in Eq.~(\ref{eq:finf})),
associated with vortex discreteness and the Magnus ``energy'' (first
term in Eq.~(\ref{eq:finf})).  For a nonuniform $\rho_s(r)$ the former
is lowered by shifting vortices out to the condensate edge (along
$-\grad\rho_s$), where $\rho_s(r)$ and the associated kinetic energy
cost is smaller. This is balanced by the Magnus force (which seeks to minimize
the distortion $\bu(\br)$) that is proportional to the difference between the
vortex velocity $\Omega\zh\times\br$ and the local superfluid
velocity $\bv_s(\br) =\Omega [ \zh \times \br - 2\zh \times
\bu_L(\br)]$. Since for a weak distortion the former energy is linear
and the latter is quadratic in $\bu(\br)$, a nontrivial vortex lattice
distortion, given in Eq.~(\ref{eq:uresult}), is always induced.

Using our main result for $\bu(\br)$, Eq.~(\ref{eq:uresult}), the superfluid
velocity can be easily computed using Eq.~(\ref{eq:ve}), giving
\be
\label{fineom2}
\bv_s(\br)\simeq   \Omega \zh \times \Big[\br
+  \frac{1}{4\omega}\big( \ln \frac{c}{\xi^2 \omega}\big)
\grad \ln \rho_s\Big],
\ee
where for simplicity we made an approximation $\grad \cdot \bu \approx
0$ inside the logarithm of Eq.~(\ref{eq:uresult}).  Since typically
$\rho_s(r)$ is concave, largest at the center of trap and decreasing
monotonically with radius (although by tailoring a trap potential it 
can be made interestingly nonmonotonic; see Sec.~\ref{SEC:examples} below),
the superfluid velocity Eq.~(\ref{fineom2}) is in fact {\em lower}
than the rigid-body result by an amount that generically increases
with radius.  Thus, in the rotating (vortex lattice) frame the
superfluid velocity is in the direction {\it opposite\/} to that of
the imposed rotation (see Fig.~\ref{fig:vplot}).

Furthermore we note that, as illustrated in Fig.~\ref{fig:vplot}, $\bv_s(\br)$ in
Eq.~(\ref{fineom2}) exhibits radial shearing, namely the superfluid
rotates with an $r$-dependent angular velocity and {\em not} simply as
a rigid body with $\bv_s(\br)=\Omega' \zh \times \br$ and $\Omega'<
\Omega$ (except in the case of Gaussian $\rho_s(r)$, see below).

A conventional fluid exhibiting such a radial shear past a
conventional (e.g., colloidal) crystal would necessarily exert a
viscous shear stress on the crystal, thereby inducing a helical-like
shear strain $\partial_r u_\phi$ ($u_\phi$ an azimuthal lattice
distortion) in the crystal. By symmetry arguments one would expect a
similar helical (azimuthal radial-shear) distortion of the vortex
lattice due to the shearing superfluid, in the direction opposite to
the fluid flow. However, within our over-simplified ideal $T=0$
superfluid analysis (no normal, quasi-particle fluid), the forces on
the vortex are purely radial (perpendicular to the superflow), and
therefore only induce a radial (longitudinal) distortion,
Eq.~(\ref{eq:uresult}).  Whether this intriguing, symmetry-suggested
helical distortion of the vortex lattice will emerge from a more
realistic (e.g., two-fluid model that includes thermally excited
quasiparticles) calculation remains an interesting open question.

Using Eq.~(\ref{eq:uresult}) together with Eq.~(\ref{eq:nvu}) (i.e., taking 
the divergence of
$\bu(\br)$), or equivalently using Eq.~(\ref{eq:ve}) (i.e., taking the curl of
$\bv_s(\br)$) we can compute the corresponding coarse-grained vortex
density by solving the differential equation that $\bar{n}_v(r)$ satisfies:
\be
\label{fineom}
\bar{n}_v(r)  = \frac{\omega}{\pi} + \frac{1}{8\pi } \grad \Big(
\frac{1}{\rho_s(r)} \grad 
\big[ \rho_s(r)\ln \frac{c}{\pi \xi^2 \bar{n}_v(r)}\big] \Big).
\ee
The physical picture embodied by Eq.~(\ref{fineom}) is straightforward
to understand.  The first term is the usual rigid body density
discussed in the Introduction, corresponding to the vortex density in the
limit of an infinite and homogeneous superfluid.  Indeed, for uniform
$\rho_s(r)$, Eq.~(\ref{fineom}) is {\it exactly} solved by the
rigid-body vortex density (i.e.~$n_{v0} = \omega/\pi$), in 
agreement with Tkachenko's
results~\cite{Tkachenko}.  The correction to $n_{v0}$ (second term
above) also vanishes in the dense vortex (or high-rotation rate, $\bar{n}_v
\xi^2 \approx 1$) limit~\cite{commentQHE}, as expected from the
discussion in Sec.~\ref{SEC:rigid}.  Hence, consistent with
experiments~\cite{Madison,Aboshaeer,Haljan,Engels,Dalibard,JILA}, an approximately uniform rigid-body  
rotation corresponding to a constant vortex density $n_{v0}$ is
predicted even in the case of a spatially-varying superfluid density,
$\rho_s(r)$.  Consistent with earlier observations, since the
condensate density variation in a trap is expected to satisfy
$\frac{\rho_s'(x)}{\rho_s(x)} \leq 0$, Eq.~(\ref{fineom}) predicts that the
coarse-grained vortex density is expected to be lower than the rigid
body result $n_{v0}$ by an amount that vanishes for a
uniform superfluid.

For smooth variations in $\rho_s(r)$, we can replace $\bar{n}_v(r)$
in the logarithm by its approximate value $\omega/\pi$, yielding our
main result
\bse
\bea
\label{eq:denslappre}
\bar{n}_v(r)  &\approx& \frac{\omega}{\pi} + \frac{1}{8\pi } \grad \Big(
\frac{1}{\rho_s(r)} \grad 
\big[ \rho_s(r)\ln \frac{c}{\xi^2 \omega}\big] \Big),
\\
&\approx& \frac{\omega}{\pi} +  \overline{c}\nabla^2 \ln \rho_s(r),
\label{eq:denslap}
\eea
\ese
that relates vortex density to superfluid density, with
\be 
\overline{c}\equiv\frac{1}{8\pi }\ln \frac{c}{\xi^2\omega}.
\ee
Because, as discussed in Sec.~\ref{SEC:free}, in the TF limit $\rho_s(r)$ is
simply determined by the trap profile with 
\be
\rho_s(r)\approx (\mu - V(r) +\frac{1}{2}m\Omega^2r^2)/g,
\ee
Eq.~(\ref{eq:denslappre}) gives an explicit prediction for the vortex density
distribution in a rotating, inhomogeneous superfluid. 

It is remarkable that, in the complementary lowest Landau level (LLL)
limit, an identical relation (with the exception of the
coefficient $\overline{c}$, which in the LLL regime is a pure number,
independent of $\xi$) emerges and was used by Ho\cite{Ho} to argue (unfortunately
incorrectly~\cite{JILA}, based on the observed approximate
uniformity of the vortex lattice) for the Gaussian form of the
condensate profile, $\rho_s(r)$ in the LLL limit.  In the recent work
by Watanabe et al.~\cite{Watanabe}, a TF profile was found to be 
the optimal one in the LLL limit (in agreement with experiments~\cite{JILA})
leading these authors to assert (consistent
with our earlier prediction~\cite{PRL}) that the vortex lattice
is nonuniform in this regime.
These latter findings were also supported by recent
numerical solutions of the problem in the LLL
regime~\cite{Cooper,MacDonald}.

\subsection{Application to specific superfluid density profiles}
\label{SEC:harmonic}

We now apply our result for $\bar{n}_v(r)$, Eq.~(\ref{fineom}), to a variety
of condensate profiles $\rho_s(r)$, realizable by tailoring the shape
of the trap potential. 

\subsubsection{Gaussian $\rho_s(r)$}

We first note that for the case of a Gaussian condensate profile
Eq.~(\ref{eq:gaussian}), the differential equation for $\bar{n}_v(r)$,
Eq.~(\ref{fineom}), is solved by a {\it uniform} density
$\bar{n}_{v,\rG}$ given by
\bse
\bea
\label{eq:gaussres}
\bar{n}_{v,G} &=& \frac{\omega}{\pi} - \frac{1}{4\pi \rnought^2} 
\ln \frac{c}{\pi\xi^2 \bar{n}_{v,\rG}},
\\
&\approx& \frac{\omega}{\pi} - \frac{1}{4\pi \rnought^2} 
\ln \frac{c}{\xi^2 \omega}.
\label{eq:gaussres2}
\eea
\ese
This solution represents a perfect hexagonal lattice, static in the
rotating frame, with a lattice constant slightly larger than the
rigid-body result.  The corresponding superfluid velocity is given by
the rigid-body form
\be
\bv_s(\br) \approx \Omega' \zh \times \br,
\ee
with
\be
\Omega' = \Omega\Big(1- \frac{1}{4\omega R^2} \ln\frac{c}{\xi^2 \omega}\Big),
\ee
slightly smaller than the imposed rotation frequency $\Omega$. This
result, (derived within London approximation) is consistent with the
prediction by Ho\cite{Ho} (discussed above) 
of a Gaussian condensate profile for a
uniform vortex lattice, derived in the complementary LLL regime.
Because of the close similarity of this vortex configuration and the
associated superflow to those of the rigid-body result, we expect
difficulties for a direct experimental verification of our prediction
in Eq.~(\ref{eq:gaussres}) for the Gaussian condensate profile.

\subsubsection{Inverted parabola $\rho_s(r)$ (harmonic trap)}

\begin{figure}[bth]
\vspace{1.4cm}
\centering
\setlength{\unitlength}{1mm}
\begin{picture}(40,40)(0,0)
\put(-50,0){\begin{picture}(0,0)(0,0)
\includegraphics{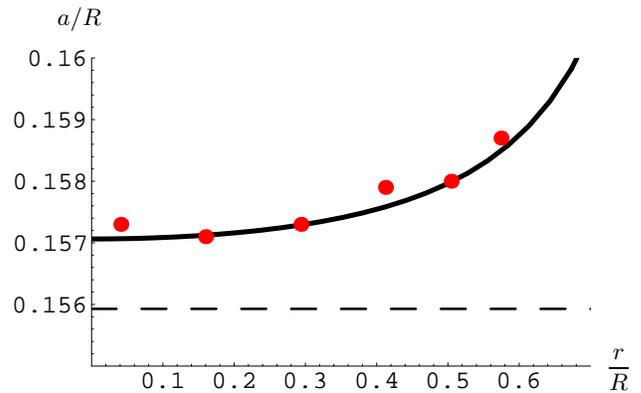}
\end{picture}}
\put(-13,51) {$a/R$}
\put(60,4.5) {$\displaystyle \frac{r}{R}$}
\end{picture}
\vspace{-.5cm}
\caption{(Color online) 
  Plot of $a/R$ (lattice spacing normalized to TF radius) as a
  function of radius using parameters $\Omega/\Omega_t = .86$ and $R =
  49 \mu m$.  Points are data adapted from Ref.~\onlinecite{JILA}.
  The dashed line is the rigid-body value of $a/R$.}
\label{fig:TFlattice1}
\end{figure}

For the most experimentally relevant case of a harmonic trap, (which,
within the TF approximation, has the condensate profile $\rho_s(r) \propto
(R^2 - r^2)$ [c.f. Eq.~(\ref{eq:tfrho})]), the vortex density profile
Eq.~(\ref{eq:denslappre}) reduces to the result advertised in 
Sec.~\ref{SEC:Intro}~\cite{commentTF}:
\be
\label{eq:tf}
\bar{n}_v(r) \simeq \frac{\omega}{\pi} - \frac{1}{2\pi}
\frac{\rtf^2}{(\rtf^2 - r^2 )^2} \ln \frac{c}{\xi^2 \omega}. 
\ee
In Fig.~\ref{fig:TFplot} we plot Eq.~(\ref{eq:tf}) for experimentally
realistic parameters $\Omega = .86 \Omega_t$, $R = 49 \mu m$ (top
solid curve), $\Omega = .57 \Omega_t$, $R = 31 \mu m$ (middle solid
curve), $\Omega = .40 \Omega_t$, $R = 25 \mu m$ (bottom solid curve)
along with the rigid-body result (dashed curve) for the case of
$^{87}$Rb.  Here, $\xi$ is taken to be the TF value~\cite{Baym96,Feder01}
\be
\xi =
\frac{\hbar}{m \Omega_t R}, \label{eq:tfxi}
\ee
and $\Omega_t
= 52 s^{-1}$.  As illustrated in Fig.~\ref{fig:TFlattice1} our prediction for
$\bar{n}_v(r)$ compares well with the recent JILA data\cite{JILA}.  
To obtain this fit, we
used values for $\Omega$ and $R$ reported in Ref.~\onlinecite{JILA},
adjusting them (from $\Omega/\Omega_t=.89$ and $R = 50 \mu m$) within
the reported error bars ($\pm 3$ and $\pm 1 \mu m$, respectively) to
find the best fit.
\begin{figure}[bth]
\vspace{1.4cm}
\centering
\setlength{\unitlength}{1mm}
\begin{picture}(40,40)(0,0)
\put(-50,0){\begin{picture}(0,0)(0,0)
\includegraphics{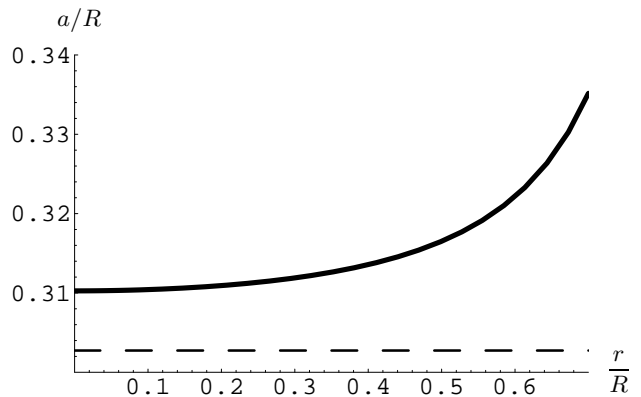}
\end{picture}}
\put(-13,51) {$a/R$}
\put(60,4.5) {$\displaystyle \frac{r}{R}$}
\end{picture}
\vspace{-.5cm}
\caption{Plot of $a/R$ (lattice spacing normalized to TF radius)  
  as a function of radius using parameters $\Omega/\Omega_t = .57$ and
  $R = 31 \mu m$.  The dashed line is the rigid-body value of $a/R$.
}
\label{fig:TFlattice2}
\end{figure}

Equivalently, the above result predicts a  hexagonal lattice constant ($a(r) =
(2/\sqrt{3} \bar{n}_v(r))^{1/2}$) that increases with $r$, as illustrated in
Figs.~\ref{fig:TFlattice1} and
\ref{fig:TFlattice2}, where in the former we again compare
our theory with the JILA experiments (with the same data as in Fig.~\ref{fig:TFplot}). 

\subsubsection{Nonmonotonic condensate profiles}
\label{SEC:examples}

It is clear that if the only variation of the condensate density is
on the scale of the trap, Eq.~(\ref{eq:uresult}) predicts a
generically small (though observable~\cite{JILA}) distortion of a
uniform hexagonal vortex lattice that vanishes as
$\nabla^2\ln\rho_s(r)\sim 1/R^2$ (for $r\ll R$).  To the extent that
observed lattices (see, e.g., Refs.~\onlinecite{Madison,Aboshaeer,Haljan,Engels,Dalibard,JILA}) are
remarkably uniform, this is a required success of our theory.
However, to more stringently test our predictions, in 
the present section we calculate 
vortex lattice distortions for nonmonotonic (but still uniaxial)
condensate density profiles $\rho_s(r)$ that also vary on a scale smaller
than the condensate's overall size $R$. We expect that such
$\rho_s(r)$ can be experimentally realized by confining a condensate
inside a nonmonotonic trap potential $V(r)$ tailored by adjusting a
combination of magnetic and optical fields.

For simplicity, ignoring the weak variation of the vortex density
inside the argument of the logarithm, i.e., replacing $1- \grad \cdot
\bu$ by unity, the distortion away from a hexagonal rigid-body array 
(which is a hexagonal lattice with lattice parameter 
$(2/\sqrt{3} n_{v0})^{1/2}$)
is given by
\be
\bu(\br) \simeq -\frac{\grad \rho_s(r)}{8 \omega \rho_s(r) }
 \ln \frac{c}{\xi^2 \omega}.
\label{eq:uresultexamples}
\ee
%
\begin{figure}[bth]
\vspace{1.3cm}
\centering
\setlength{\unitlength}{1mm}
\begin{picture}(40,40)(0,0)
\put(-50,0){\begin{picture}(0,0)(0,0)
\includegraphics{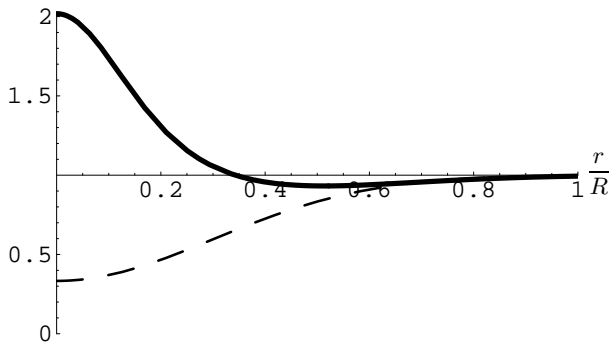}
\end{picture}}
\put(58,24.5) {$\displaystyle \frac{r}{R}$}
\end{picture}
\vspace{-.5cm}
\caption{Main: The solid line depicts the vortex density $\bar{n}_v(r)/n_{v0}$  as a function of radius for the 
  condensate profile Eq.~(\ref{eq:profile1}).  The dashed line plots $\rho_s(r)/\rho_0$.
}
\label{fig:distorta}
\end{figure}
\begin{figure}
%
 \epsfxsize=\columnwidth
%
\centerline{\epsfbox{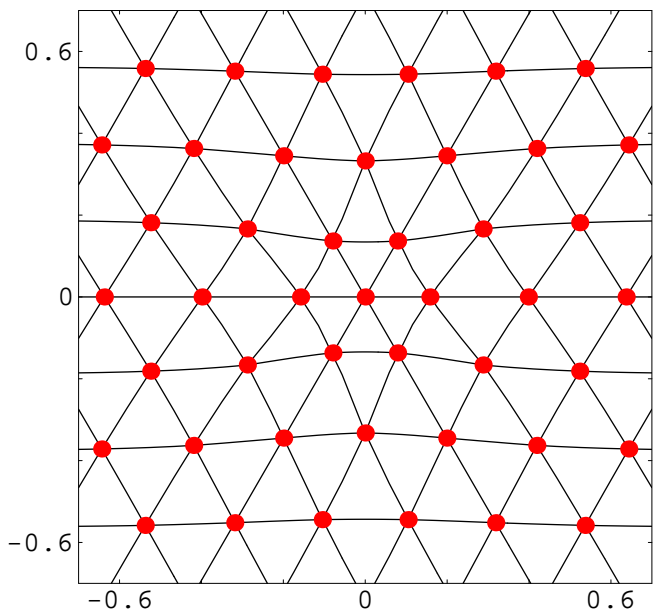}}
\caption{(Color online) Distorted vortex lattice (points) arising from the condensate profile 
Eq.~(\ref{eq:profile1}) containing
a depletion near the origin with lines to guide the eye. 
}
\label{fig:distort}
\end{figure}

We first consider perhaps the simplest nonmonotonic condensate profile
with a dip in its center of spatial extent $\ell$, that we model by
\be
\rho_s(r) = \rho_0 - \rho_1 \exp(-r^2/2\ell^2),  
\label{eq:profile1}
\ee
with $\rho_1 <\rho_0$, where the finite asymptotic density $\rho_0$ is
strictly speaking not a constant but varies on a much larger scale $R$
and is determined by an overall large-scale trap with $R\gg\ell$.  A
qualitatively similar condensate profile can be generated by a
``Mexican hat''-like trap potential, as experimentally demonstrated by
the research groups of Dalibard~\cite{Dalibard} and
Ketterle~\cite{Ketterle}.
Using such $\rho_s(r)$ inside Eq.~(\ref{eq:uresult}) and Eq.~(\ref{fineom}) we find
\bse
\bea
&&\bu(\br)= -\frac{1}{8\omega \ell^2}\ln\frac{c}{\xi^2 \omega}
\frac{\br}{\frac{\rho_0}{\rho_1}\exp(\frac{r^2}{2\ell^2}) - 1},
\label{eq:ugauss}
\\
&&\bar{n}_v(r)= \frac{\omega}{\pi} + \frac{1}{8\pi \ell^2}\ln\frac{c}{\xi^2 \omega}
\Big[\frac{2}{\frac{\rho_0}{\rho_1}{\rm e}^{r^2/2\ell^2}-1}
\nonumber \\
&&\qquad\qquad
- \frac{r^2}{\ell^2}
\frac{\frac{\rho_0}{\rho_1}{\rm e}^{r^2/2\ell^2}}
{(\frac{\rho_0}{\rho_1}{\rm e}^{r^2/2\ell^2}-1)^2}
\Big].
\label{eq:nvgauss}
\eea
\ese
We plot the corresponding vortex density Eq.~(\ref{eq:nvgauss}) 
in Fig.~\ref{fig:distorta} for 
a large slowly-rotating condensate of $^{87}$Rb atoms with parameters
($R = 500 \mu m$, $\Omega = 0.2 s^{-1}$, $\rho_1/\rho_0 = 0.67$, and
$\ell = 0.3 R$) that are slightly different than typical values of
present-day experiments to enhance visualization of the distortion.
Consistent with our earlier qualitative discussion, $\bar{n}_v(r)$ is
largest where $\rho_s(r)$ is smallest. Near $r = .5R$,
$\bar{n}_v(r)$ actually drops below its rigid-body value.  The
corresponding distortion $\bu(\br)$ of the vortex lattice is illustrated
in Fig.~\ref{fig:distort}, showing vortices displacing against the
local gradient in $\rho_s(r)$.
\begin{figure}[bth]
\vspace{1.4cm}
\centering
\setlength{\unitlength}{1mm}
\begin{picture}(40,40)(0,0)
\put(-50,0){\begin{picture}(0,0)(0,0)
\includegraphics{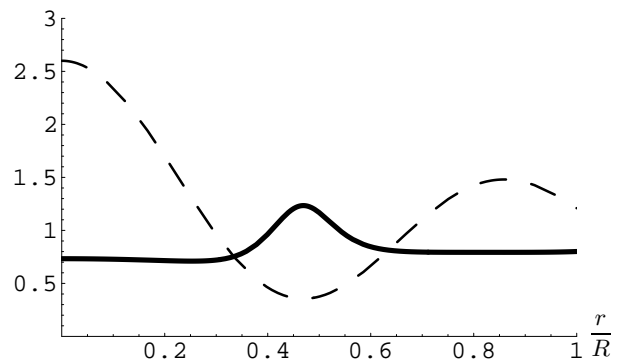}
\end{picture}}
\put(58,4.5) {$\displaystyle \frac{r}{R}$}
\end{picture}
\vspace{-.5cm}
\caption{Plot of condensate profile $\rho_s(r)/\rho_0$ (dashed line)
  and vortex density $\bar{n}_v(r)/n_{v0}$ (solid line) for the
  oscillatory profile Eq.~(\ref{eq:oscbessel}) with parameters $R = 49
  \mu m$ and $\ell = 6 \mu m$.  }
\label{fig:osca}
\end{figure}
\begin{figure}
%
 \epsfxsize=\columnwidth
%
\centerline{\epsfbox{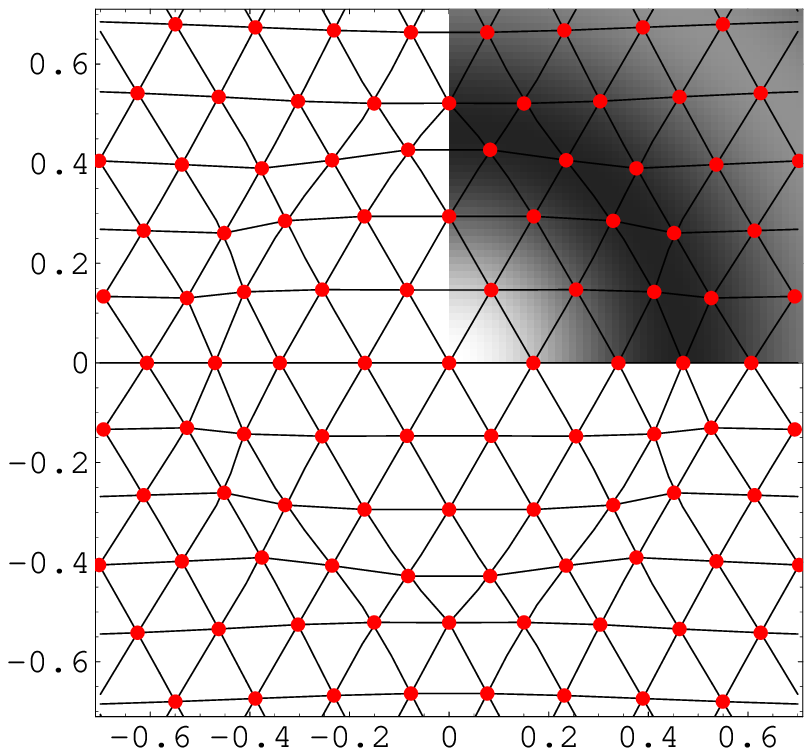}}
\caption{(Color online) Vortex positions (measured relative to $R$) 
  for the profile Eq.~(\ref{eq:oscbessel}) with lines to guide the
  eye.  The parameters $R = 49 \mu m$ and $\ell = 6 \mu m$. For the
  upper right quadrant, we have shown a density plot of $\rho_s(r)$
  with the light areas denoting regions of high $\rho_s$.  }
\label{fig:osc}
\end{figure}
\begin{figure}[bth]
\vspace{1.4cm}
\centering
\setlength{\unitlength}{1mm}
\begin{picture}(40,40)(0,0)
\put(-50,0){\begin{picture}(0,0)(0,0)
\includegraphics{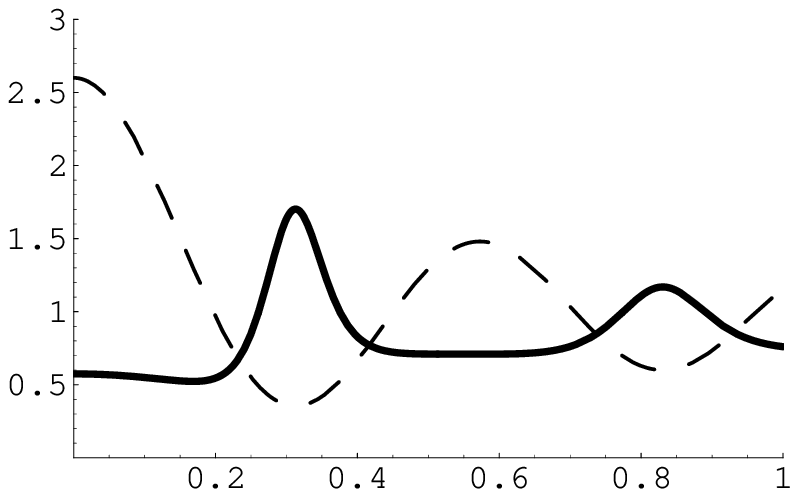}
\end{picture}}
\put(58,4.5) {$\displaystyle \frac{r}{R}$}
\end{picture}
\vspace{-.5cm}
\caption{Plot of condensate profile $\rho_s(r)/\rho_0$ (dashed line)
  and vortex density $\bar{n}_v(r)/n_{v0}$ (solid line) for the
  oscillatory profile Eq.~(\ref{eq:oscbessel}) with parameters $R = 49
  \mu m$ and $\ell = 4 \mu m$.  }
\label{fig:osca2}
\end{figure}
\begin{figure}
%
 \epsfxsize=\columnwidth
\centerline{\epsfbox{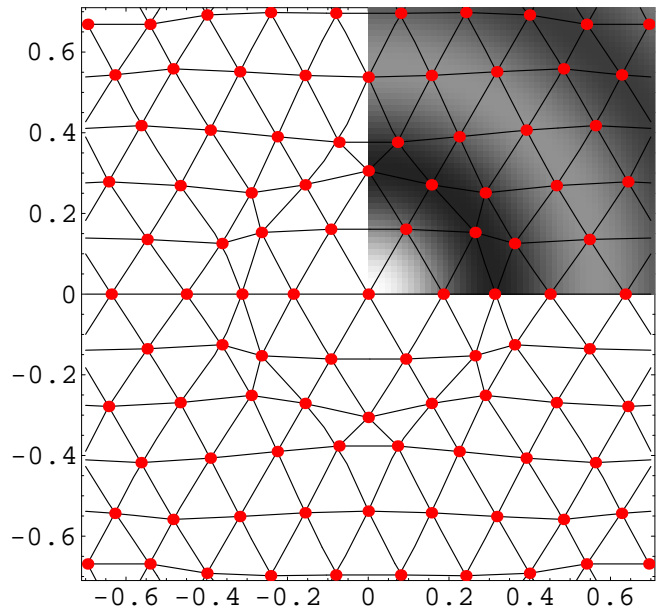}}
\caption{(Color online) Vortex positions (measured relative to $R$) 
  for the profile Eq.~(\ref{eq:oscbessel}) with lines to guide the
  eye.  The parameters $R = 49 \mu m$ and $\ell = 4 \mu m$.  For the
  upper right quadrant, we have shown a density plot of $\rho_s(r)$
  with the light areas denoting regions of high $\rho_s$.  }
\label{fig:osc2}
\end{figure}

Since $\bu(\br)$ is constrained to vanish at the trap center where
$\grad\rho_s=0$ (see Eq.~(\ref{eq:uresultexamples})), to maximize the distortions
associated with a spatially-varying $\rho_s(r)$ it makes sense to
consider a condensate profile that exhibits its most rapid variation
away from this symmetry point, i.e., at nonzero radii.  Motivated by
this we consider a trap potential and therefore a condensate density
that oscillates with $r$.  For concreteness we model such $\rho_s(r)$
with a Bessel function 
\bea
\label{eq:oscbessel}
\rho_s(r) = \rho_0 + \rho_1 J_0(r/\ell),
\eea
with $\ell$ characterizing the length-scale of the oscillations and
$\rho_0$ (as in the previous example) weakly $r$ dependent on a much
larger overall trap scale $R$, with $R\gg\ell$.  To keep the overall
condensate density positive we choose $\rho_1 < 2.48 \rho_0$.  For
this condensate profile (displayed in Fig.~\ref{fig:osca}, dashed
curve) we compute the resulting vortex density $\bar{n}_v(r)$ and illustrate
it in Fig.~\ref{fig:osca} (solid curve) for parameters (TF radius $R =
49 \mu m$, trap frequency $\Omega_t = 52 s^{-1}$, $\ell = 6 \mu m$,
$\rho_1/\rho_0 = 1.6$, $\Omega/\Omega_t = .86$ and with $\xi$ given by
Eq.~(\ref{eq:tfxi})) that are consistent with recent Rb experiments
at JILA~\cite{Engels,JILA}.  At the location of the dip in
the condensate density ($r \approx 0.5R$), $\bar{n}_v(r)$ exhibits a corresponding
maximum.  The effect on the locations of the vortices is quite
striking, as shown in Fig.~\ref{fig:osc}: the vortex rings near the
dip are compressed together.  In the upper right
quadrant we have included a density plot of $\rho_s(r)$ to show how
regions of low $\rho_s(r)$ (which are darker) induce higher local
vortex density. Thus, this profile nicely illustrates the principle
that vortices migrate to regions where $\rho_s$ is small to lower the kinetic
energy.

The case of a smaller value of $\ell$ ($\ell = 4\mu m$), such that
$\rho_s(r)$ has two minima before $r=R$ is reached, is illustrated in
Figs.~\ref{fig:osca2} and ~\ref{fig:osc2}, again demonstrating how (as expected) vortices
congregate near radial minima of $\rho_s(r)$, with a local density and
superfluid velocity that actually {\it exceed\/} their rigid-body values.  

\subsection{Effect of inhomogeneous vortex density on condensate profile}

Thus far, to calculate the average vortex density $\bar{n}_v(r)$ we have used the TF
expression for the condensate density $\rho_s(\br)$,
Eq.~(\ref{eq:tfrhopre}), that was derived in Sec.~\ref{SEC:free} within an
approximation of a rigid-body superflow profile. However, as we have
shown in the previous section, the superflow profile is itself
modified inside an inhomogeneous condensate. Hence in principle we
need to determine $\bar{n}_v(\br)$ and $\rho_s(\br)$ by self-consistently
solving the coupled equations Eq.~(\ref{eq:rhoeom}) and Eq.~(\ref{fineom}), together
with $\bar{n}_v(\br)=\frac{m}{2\pi \hbar} \grad\times\bv_s$. However, as we show below, for a
smooth trap potential, to a good approximation, the effect of the
inhomogeneity of the vortex distribution on the equilibrium
$\rho_s(\br)$ is negligible, with corrections vanishing as a higher power
in $\grad\rho_s/\rho_s\sim 1/R$.  

To show this, we repeat the steps of Sec.~\ref{SEC:phasegrad}, but now
using the full bosonic energy density Eq.~(\ref{eq:f}) instead of
keeping only the $\theta$-dependent terms (i.e.~as in
Eq.~(\ref{eq:cellfree})). This gives:
\bea 
\hspace{-1cm}&&\hspace{-1cm}E\simeq \frac{\hbar^2}{2m} \int d^2 r \rho_s(\br)\Big[ \pi
\bar{n}_v \ln \frac{1}{\pi \xi^2 \bar{n}_v} + 4\omega^2 u_L(\br)^2)\Big]
\nonumber
\\
&&+ \int d^2 r \Big[ (V - \frac{1}{2} m \Omega^2 r^2
-\mu)\rho_s(\br) + \frac{g}{2} \rho_s(\br)^2\Big],
\label{eq:finalrhoenergy}
\eea
where we remind the reader that $\bar{n}_v$ is related to $\bu_L$ via
Eq.~(\ref{eq:nvu}), and, in the spirit of the TF approximation, we have
still neglected the subdominant (away from the condensate edge, $r\ll
R$) $\grad\rho_s$ contributions to the energy.  The corresponding,
coupled Euler-Lagrange equations ($\delta E/\delta u_L=0$, $\delta
E/\delta\rho_s=0$) that determine $\bar{n}_v(\br)$ (through $\bu_L$) and
$\rho_s(\br)$ are given by:
\bse 
\bea
\label{eq:ueom}
&&\bu_L(\br) = -\frac{1}{8 \omega}
\frac{\grad \rho_s(r)}{\rho_s(r)}\ln \frac{c}{\xi^2 \omega },
\\
&&\rho_s(\br) = \frac{1}{g}\big[
\mu - b\hbar \Omega - (V(r) - \frac{1}{2} m \Omega^2 r^2)\big]
\nonumber 
\\
&& \qquad 
- \frac{1}{g} \big[
2m \Omega^2 u_L^2 - \hbar \Omega \grad \cdot \bu \ln \frac{1}{\xi^2 \omega}
\big],
\label{eq:rhoeomp1}
\eea
\ese
where, as before, in Eq.~(\ref{eq:ueom}) we have replaced $1 - \grad
\cdot \bu \simeq 1$ in the argument of the logarithm of
Eq.~(\ref{eq:uresult}), and defined a parameter $b\equiv \frac{1}{2}
\ln \frac{1}{\xi^2 \omega}$. While the equation for $u_L$ remains
unchanged, the equation for $\rho_s(\br)$ is modified by the vortex kinetic
energy, that, as expected, suppresses it.

Although these coupled equations are in general quite nontrivial, for
a smooth trap the distortion $\bu(\br)$ is small, vanishing with
$\grad\rho_s/\rho_s\sim 1/R$. Hence, they can be straightforwardly solved
by iteration in this small parameter.  To lowest order we can simply
neglect the vortex lattice distortion $\bu(\br)$ inside the $\rho_s(\br)$
equation, obtaining our previous symmetric TF result
\be 
\rho_s(r) \approx \frac{1}{g}\big[ \mu - b\hbar \Omega - \frac{1}{2}
m (\Omega_t^2 - \Omega^2) r^2\big],
\label{eq:rhoeomp2}
\ee
corrected only by $b\hbar \Omega$, that is small compared to $\mu$.
Higher order corrections to $\rho_s(r)$ are easily obtained by an
iterative proceedure, thereby generating a perturbative expansion in
$1/R$. Hence, in the high rotation limit, $\Omega \to
\Omega_t$ (corresponding to a divergent $R(\Omega)$, Eq.~(\ref{eq:romega})) the TF
condensate density profile Eq.~(\ref{eq:tfrhopre}) is an  
accurate description, in agreement with recent results obtained within the
lowest Landau level approximation~\cite{Watanabe,Cooper}.

\section{Elastic theory and Tkachenko modes}
\label{SEC:elastic}

Having determined the equilibrium vortex distribution in a spatially
inhomogeneous rotating condensate, we now study long-wavelength vortex
fluctuations about this  slightly inhomogeneous equilibrium vortex
state.  To this end we derive a vortex lattice elastic energy by
expanding the total energy Eq.~(\ref{eq:finf}) to harmonic order in
small distortions $\beps(\br)$ about the ground-state configuration
$\bu_0(\br)$ [satisfying Eq.~(\ref{eq:uresult})], defined by $\bu(\br) =
\bu_0(\br) + \beps(\br)$.  We find
\bse 
\bea \hspace{-.5cm}E_{el}[\beps(\br)] \!\!&=& \!\frac{\hbar^2}{2m}\!\! \int d^2 r
\rho_s(r)\Big[ 4 \omega^2 \epsilon_L^2\! -\! \frac{\omega}{2} \frac{(\grad
  \cdot \beps)^2}{1\!-\!\grad \cdot \bu_0} \Big],
\label{eq:elen1} 
\\
&\simeq&  \!\frac{\hbar^2}{2m}\!\! \int d^2 r  \rho_s(r)\Big[ 4 \omega^2 \epsilon_L^2 - \frac{\omega}{2}
(\grad \cdot \beps)^2 \Big],
\label{eq:elen2}
\eea
\ese
where in Eq.~(\ref{eq:elen2}) we have approximated $\bu_0 \simeq 0$.

This compressional energy must be augmented by the elastic energy of
the vortex lattice {\it shear\/} deformation, that is clearly missed
at this level of a coarse-grained (density-functional) approximation.
We follow Baym and Chandler~\cite{baym1} and fix the shear modulus using
Tkachenko's~\cite{Tkachenko,Tkachenko2} exact result for the uniform
condensate, that we expect (given the high uniformity of the vortex
lattice observed in experiments and demonstrated here analytically) to
be a good approximation even for our case of a spatially-varying
$\rho_s(r)$.  This yields
\bea && E_{el} \simeq \frac{\hbar^2}{2m} \int d^2 r\rho_s(r)\Big[ 4
\omega^2 \epsilon_L^2 - \frac{\omega}{2} (\grad \cdot \beps)^2
\nonumber
\\
&& \qquad +\frac{\omega}{4} \Big( \frac{\partial \epsilon_x}{\partial
  x} -\frac{\partial \epsilon_y}{\partial y} \Big)^2 +\frac{\omega}{4}
\Big( \frac{\partial \epsilon_x}{\partial y} +\frac{\partial
  \epsilon_y}{\partial x} \Big)^2 \Big].
\label{eq:elenfin}
\eea

The dynamics of long-wavelength elastic vortex fluctuations is
governed by the balance of the Magnus ``force'' against the
``elastic'' force associated with the distortion $\beps(\br,t)$ of the
vortex lattice
\be -\frac{\pi}{\omega} \frac{\delta E_{el}}{\delta \beps(\br,t)} +
2\pi\hbar\rho_s\zh \times \dot{\beps}(\br,t) = 0,
\label{eq:tkacheom}
\ee
where $\dot{\beps} \equiv \frac{d\beps}{dt}$ and the inverse of the
average vortex density factor $1/n_{v0}=\pi/\omega$ arises from the
Jacobian relating the variation with respect to $\br_i$ to the
functional variation with respect to $\beps(\br_i)$.  Carrying out the
functional derivative yields the following equation for
$\beps(\br,t)$:
\begin{widetext}
\bea
\label{eq:tmodebig}
\rho_s \zh \times \dot{\beps} =  -\frac{2 \hbar \omega}{m} \grad\!\!\int d^2 \br' \rho_s(r') 
\beps_{L}(\br',t) \!\cdot \! \grad G(\br' \!-\! \br)\! -\! \frac{\hbar}{8m} \big(
[\grad \rho_s \!\cdot\! \grad] \beps\! +\! [ \zh \cdot (\grad \rho_s \!\times\! \grad) ] \zh \times \beps
+\rho_s \nabla^2 \beps \!-\! 2\grad[\rho_s \grad \!\cdot\! \beps]
\big).
\eea
\end{widetext}
A full solution of Eq.~(\ref{eq:tmodebig}) is beyond the scope of this
paper and we leave it for future research.  In spite of its complexity
it is reassuring that for a uniform condensate ($r$-independent
$\rho_s$) the divergence and curl of Eq.~(\ref{eq:tmodebig}) are equivalent to
Eqs.~(71) and ~(72) of Ref.~\onlinecite{baym1}, respectively.  In the
following, we proceed by making a simple analytic approximation to
Eq.~(\ref{eq:tmodebig}) to bring out the leading order effect of the
condensate inhomogeneity on the Tkachenko waves.

For long-wavelength Tkachenko modes, with $\frac{1}{a} \gg k \agt
\frac{1}{R}$, it is clear that the first term on the right side of
Eq.~(\ref{eq:tmodebig}) (which is purely longitudinal) dominates over
the second since $\omega\propto 1/a^2$.  Taking the divergence of both
sides of Eq.~(\ref{eq:tmodebig}) and neglecting the subdominant second
term, we find
\be
\grad \times \dot{\beps} = -2\Omega \grad \cdot \beps \zh ,
\label{eq:connection}
\ee
to leading order in a double expansion in spatial derivatives of
$\beps$ and $\rho_s$.  The transverse part of Eq.~(\ref{eq:tmodebig}) is
obtained by taking its curl, which leads to
\bea
&&\hspace{-.5cm}\grad \times[ \rho_s \zh \times \dot{\beps}]
= -\frac{\hbar}{8m} \Big[ \grad \times (\rho_s \nabla^2 \beps) 
\nonumber \\
&&+\grad \times \Big(
[\grad \rho_s \cdot \grad] \beps + [ \zh \cdot (\grad \rho_s \times \grad) ] \zh \times \beps \Big)    
\Big].
\label{eq:ttrans}
\eea
Henceforth, for simplicity, we restrict our attention to the case of a
Gaussian $\rho_s(r)$, Eq.~(\ref{eq:gaussian}), which satisfies $\grad
\rho_s(r) = -\br\rho_s(r)/R^2$.  To leading order in spatial gradients, the
quantity in parentheses in the second line of Eq.~(\ref{eq:ttrans})
reduces to
\be
[\grad \rho_s \cdot \grad] \beps + [ \zh \cdot (\grad \rho_s \times \grad) ]
 \zh \times \beps \simeq \frac{\rho_s \beps}{R^2}.
\label{eq:ttrans2}
\ee
Inserting Eq.~(\ref{eq:ttrans2}) into Eq.~(\ref{eq:ttrans}) and 
expanding the result to leading order in derivatives of $\rho_s$, we have 
\be
\grad \cdot \dot{\beps}\zh \simeq -\frac{\hbar}{8m R^2} 
\grad \times \beps- \frac{\hbar}{8m } \nabla^2(\grad \times \beps),
\ee
where we have used $ \grad \times \nabla^2 \beps = \nabla^2(\grad
\times \beps)$.  Combining this result with Eq.~(\ref{eq:connection}),
we finally obtain an equation for the transverse vortex fluctuations
$\grad \times \beps$:
\be
\label{eq:fintrans}
\frac{d^2}{dt^2}(\grad \times \beps) = \frac{\hbar \Omega}{4 m} 
\left(\frac{1}{R^2} + \nabla^2\right)(\grad \times \beps).
\ee
The spatial and temporal Fourier transform of this linear wave equation
then immediately gives the Tkachenko mode dispersion $\Omega_T(k)$:
\be 
\Omega_T(k) = \sqrt{\hbar \Omega/4m} \sqrt{k^2 - R^{-2}}, 
\ee
that, in the uniform limit $R\to \infty$, recovers the standard linear
dispersion result $\Omega_T(k) = k \sqrt{\hbar \Omega/4m}$ (see
Refs.~\onlinecite{Tkachenko2,baym1}).  Given the small gradient in
$\rho_s(r)$ expansion that led to this result, the wavevector is
obviously limited to the physically sensible range $1/a > k > 1/R$.
No doubt a more detailed analysis of Eq.~(\ref{eq:tmodebig}) is needed to obtain a
more accurate prediction for Tkachenko eigenmodes and their dispersion
in an inhomogeneous condensate characterized by a generic condensate
profile $\rho_s(r)$. We leave such analysis, as well as an extension
of Eq.~(\ref{eq:tmodebig}) that includes condensate 
{\it density\/} fluctuations, 
as discussed in the recent work by Baym~\cite{baym03}
(see also Ref.~\onlinecite{Sonin}), to future
research.

\section{Summary and Conclusions}
\label{SEC:concluding}

To summarize, we have presented a London theory of a vortex state in an
inhomogeneous rotating Bose-Einstein condensate. Our most important
result (Eq.~(\ref{fineom})) is the relation between the vortex density $\bar{n}_v(r)$
and the superfluid density $\rho_s(r)$. When applied to a harmonic
trap, this result provides a simple explanation for the observed
highly regular vortex arrays in such strongly inhomogeneous
condensates, and the observed Thomas-Fermi parabolic condensate
profile.  This relation also predicts a slight inhomogeneity in the
vortex spatial distribution, that has recently (since our prediction)
been observed in experiments and simulations from JILA~\cite{JILA}.  
As we discussed in the
main text, this relation between $\bar{n}_v(r)$ and $\rho_s(r)$ can be more
stringently tested by studying vortex lattice distortions induced by
nonmonotonic condensate profiles, tailored with various trapping
potentials. 

As an important digression, we also studied the superflow in the
vicinity of an isolated vortex, calculating the superfluid velocity
distortion induced by a spatially-varying condensate. The associated
additional contribution to the superflow is directed orthogonal to the
displacement of the vortex away from the trap center, thereby
providing a simple explanation for precession of such vortices about
the trap center with radially-independent frequency.

With respect to our vortex lattice result, it is important to stress
that our prediction of the deviation of the vortex density from the
spatially independent (rigid-body) result is more than just a
suppression of the total vortex number (a somewhat trivial and
experimentally difficult\cite{commentOmega} effect to detect, that
follows directly from the existence of $\Omega_{c1}$, as can be seen
from arguments in Sec.~\ref{SEC:omc1} and Fig.~\ref{fig0}). 
In particular, it is not a simple surface effect,
such as, e.g., a missing vortex ring at the edge of the condensate, as for
example observed in simulations of Feder and Clark~\cite{Feder01}
and predicted theoretically for uniform superfluids some time ago~\cite{Stauffer}.

Our prediction of a radially increasing vortex lattice spacing across
the {\em bulk} of a condensate (that does not require a precise
measure of $\Omega$\cite{commentOmega}), shows quantitative agreement
with recent JILA data~\cite{JILA}, exhibiting the same $\sim 2\%$ lattice
distortion.

We conclude by noting that a number of important extensions of our
work remain. Based on symmetry considerations we have suggested that
an azimuthally directed vortex lattice radial-shear distortion
$\partial_r u_\phi$ will be induced by the radially-shearing
superfluid.  Clearly, it is important to determine theoretically
whether such interesting chiral vortex lattice distortion is indeed
present. This should be possible to assess within either a two-fluid
hydrodynamic model or by incorporating thermally-excited Bogoliubov
quasi-particles into our theory. A more careful treatment of vortex
discreteness (that goes beyond our simple density-functional-like
approximation), from which a finite vortex lattice shear modulus must
emerge is also highly desirable. This would allow a first-principles
treatment of Tkachenko modes, that is more precise than that presented
here.  Finally, the extension of our work to three dimensions, where a
nontrivial $z$-dependence of the vortex-line density profile is
expected, is another important future direction.

\acknowledgments 
We gratefully acknowledge useful discussions with A. Andreev, I.
Coddington, E. Cornell, P. Engels, A. Fetter and V. Schweikhard as well as
support from NSF DMR-0321848 and the Packard Foundation.
\appendix
\section{Vortex precession in a uniform superfluid}
\label{SEC:appendix}
Our derivation in Sec.~\ref{SEC:ind} of vortex precession in an inhomogeneous
trapped superfluid can be complemented by an exact calculation of
vortex precession in a {\em homogeneous} condensate confined to a finite
``bucket'' of radius $R$ (see Ref.~\onlinecite{McGee}). To model this system we take
$\rho_s$ be a constant $\rho_0$ for $r<R$ and zero for $r>R$.  For this
case, the single-vortex problem is solved by the method of images to
enforce the boundary condition of no outward current (i.e.~$\grad
\theta$ must be orthogonal to the boundary)\cite{commentBucket}.

It is straightforward to verify that a single vortex at $\br_0$ in the
bucket geometry has the solution
\be
\label{bcTheta}
\grad \theta(\br) = \frac{\zh \times (\br - \br_0)}{(\br - \br_0)^2} -
\frac{\zh \times (\br - \br_i)}{(\br - \br_i)^2}, \ee
where $\br_i = \br_0 R^2/r_0^2$ is the \lq\lq image charge\rq\rq.
Thus, although the condensate is homogeneous in the bulk,
the nontrivial boundary condition
necessarily introduces a curl-free (in the bulk) correction
$\grad\theta_a$ to the superflow (second term in Eq.~(\ref{bcTheta})),
that has a similar physical effect as that due to the bulk condensate
inhomogeneity studied in Sec.~\ref{SEC:ind}.

As for an inhomogeneous condensate, the background superflow at the
vortex and thus the vortex dynamics are entirely determined by the image charge
term.  Evaluating the associated contribution at $\br_0$ and using
Eq.~(\ref{eq:magnus0}) we find
\be
\dot{\br}_0 = \frac{\hbar}{m} \frac{\zh \times \br_0}{R^2 - r_0^2},
\ee
showing that a vortex located off-center in a finite ``bucket'' will
precess about the origin in a sense similar to that embodied in
Eq.~(\ref{eq:singlevortdyn}) and Eq.~(\ref{eq:singlevortdyn2}), with
the frequency near the center of order $\hbar/m R^2$, corresponding to
a unit of angular momentum for a boson at the boundary of the
condensate, as found for an inhomogeneous condensate.


\begin{thebibliography}{10}
\bibitem{onsager} L. Onsager, Nuovo Cim. Suppl. 2 {\bf 6}, 279 (1949).
%
\bibitem{feynman}
R.P. Feynman, in {\it Progress in Low Temperature Physics}, C.J. Gorter, ed. (1955).
%
\bibitem{Matthews}
%
%
M.R. Matthews, B.P. Anderson, P.C. Haljan, D.S. Hall, C.E. Wieman and E.A. Cornell,
Phys. Rev. Lett. {\bf 83}, 2498 (1999).
\bibitem{Madison}
K.W. Madison, F. Chevy V. Bretin and J. Dalibard,
Phys. Rev. Lett. {\bf 86}, 4443 (2001).
\bibitem{Aboshaeer}
J.R. Abo-Shaeer, C. Raman, J.M. Vogels and W. Ketterle,
Science {\bf 292}, 476 (2001).
%
\bibitem{Haljan}
P.C. Haljan, I. Coddington, P. Engels and E.A. Cornell,
Phys. Rev. Lett. {\bf 87}, 210403 (2001).
\bibitem{Engels}
P. Engels, I. Coddington, P.C. Haljan, V. Schweikhard and E.A. Cornell,
Phys. Rev. Lett. {\bf 90}, 170405 (2003).
\bibitem{Dalibard}
Vincent Bretin, Sabine Stock, Yannick Seurin and Jean Dalibard,
Phys. Rev. Lett {\bf 92}, 050403 (2004).
%
\bibitem{JILA}
I. Coddington, P.C. Haljan, P. Engels, V. Schweikhard, S. Tung and E.A.
Cornell, cond-mat/0405240.
\bibitem{Anglin}
J.R. Anglin and M. Crescimanno,
cond-mat/0210063.
%
\bibitem{PRL}
Daniel E. Sheehy and Leo Radzihovsky,
cond-mat/0402637.
\bibitem{Feder01}
David L. Feder and Charles W. Clark,
Phys. Rev. Lett., {\bf 87}, 190401 (2001).
\bibitem{commentSCII} In superconductors, vortex pinning by $\rho_s(r)$
  inhomogeneities is due to the lowering of the loss of condensation energy
(i.e.~an increase in condensation energy)
  when the ``normal'' vortex core spatially coincides with the material
  inhomogeneity. Because vortex motion generates electric fields
  (Josephson relation), vortex pinning is in fact crucial for type-II
  superconductivity.
\bibitem{commentRhos} For our purposes here, it is sufficient to
  confine our analysis to a zero-temperature mean-field approximation.
  At this level, for weak interactions appropriate for a dilute gas
  the condensate depletion is vanishingly small, allowing us to ignore
  the distinction between the superfluid density $\rho_s(r)$ and the
  condensate density $|\Phi(r)|^2$.
\bibitem{commentQHE} In this article we do not consider the extremely
  high rotation rate regime, where Landau level quantization
  is important, boson and vortex densities are comparable and
  therefore vortices must again be treated as discrete and furthermore
  as quantum mechanical objects. Even for a large filling fraction
  $N_b/N_v$ so that the rotated superfluid remains a well-ordered
  vortex lattice, for sufficiently weak interactions, low density and
  large rotation rate, the coherence length grows to become comparable to
  inter-vortex spacing, invalidating our London approximation.  The
  many-body ground state is then still well-described by a product of
  single-particle vortex-lattice wavefunctions (all bosons occupying
  this same vortex state), but (unlike in London approximation) in this
  so-called lowest Landau level (LLL) regime, constructed as a superposition
  of orbitals constrained to the LLL. As argued theoretically~\cite{BaymPethick}
  and demonstrated convincingly experimentally~\cite{JILA},
  within this high rotation LLL regime, the
  $\omega\xi^2\propto(\xi/a)^2$ saturates, approaching a constant.
\bibitem{BaymPethick} 
Gordon Baym and C.J. Pethick
cond-mat/0308325.
\bibitem{Tkachenko}
V.K. Tkachenko, 
Zh. Eksp. Teor. Fiz. {\bf 49}, 1875 (1965) 
[Sov. Phys. JETP {\bf 22}, 1282 (1966)].
\bibitem{Watanabe}
Gentaro Watanabe, Gordon Baym and C.J. Pethick,
cond-mat/0403407.
\bibitem{Cooper}
N.R. Cooper, S. Komineas and N. Read, 
cond-mat/0404112.
\bibitem{Pismen}
B.Y. Rubinstein and L.M. Pismen, 
Physica D {\bf 78}, 1 (1994); see also L.M. Pismen {\it Vortices 
in Nonlinear fields}, Oxford (1999).
%
%
\bibitem{Fedichev}
P.O. Fedichev and G.V. Shlyapnikov,
Phys. Rev. A {\bf 60}, 1779 (1999).
%
%
\bibitem{Linn}
%
Marion Linn and Alexander L. Fetter, Phys. Rev. A {\bf 61}, 063603 (2000).
\bibitem{Lundh}
Emil Lundh and P. Ao,
Phys. Rev. A {\bf 61}, 063612 (2000).
\bibitem{McGee}
S.A. McGee and M.J. Holland,
Phys. Rev. A, {\bf 63}, 043608 (2001).
%
\bibitem{Svidzinsky} 
Anatoly A. Svidzinsky and Alexander L. Fetter, Phys. Rev. Lett. {\bf 84}, 5919 (2000);
Phys. Rev. A {\bf 62}, 063617 (2000).
\bibitem{Anglin02}
J.R. Anglin, 
Phys. Rev. A {\bf 65}, 063611 (2002).
\bibitem{Svidzinskynote}
The result for the superfluid velocity near an
  off-axis vortex, that we find here, was first published in 
Ref.~(\onlinecite{Pismen}) and generalized to the three-dimensional 
case in Ref.~(\onlinecite{Svidzinsky}).  
  However, to the best of our knowledge, the extension of the solution for
  ${\bf v}_s(r)$ to the region outside the immediate vicinity of the vortex
  has not appeared in the literature until our work\cite{PRL}.
\bibitem{Anderson}
B.P. Anderson, P.C. Haljan, C.E. Wieman and E.A. Cornell,
Phys. Rev. Lett. {\bf 85}, 2857 (2000).
%
\bibitem{mftCoherentZ} The energy density for the Bose condensate can
  be straightforwardly obtained from a standard interacting boson
  Hamiltonian, with boson operators approximated by a classical
  condensate field, appropriate for the BEC state. Equivalently, the
  same result can be obtained from a mean-field approximation to the
  coherent state path integral for the partition function for
  the interacting boson problem.
\bibitem{commentTF} This is what is known as the Thomas-Fermi
  approximation, which ignores the kinetic energy (i.e.~gradients in the
  condensate wavefunction) in comparison to the trap potential and
  boson interactions and which is valid away from the condensate edges. Not
  surprisingly, this neglect of kinetic energy leads to unphysical
  anomalies near the condensate edges such as divergent corrections to the
  vortex density $\bar{n}_v(r)$ as $r \to \rtf$. Although in a more
  careful approximation these divergences do not appear, we expect
  that our small gradient (in $\ln \rho_s(r)$) analysis breaks down near
  the edge of the condensate, where the corrections to the uniform
  vortex density and rigid-body superfluid velocity are large.
\bibitem{leggett}
See, e.g., A.J. Leggett, Rev. Mod. Phys. {\bf 71}, S318 (1999).
%
\bibitem{Baym96} 
Gordon Baym and C.J. Pethick, 
Phys. Rev. Lett. {\bf 76}, 6 (1996).
%
\bibitem{commentDiscreteN} We must remark that, since $N$ is a
  discrete variable, in actuality one expects steps in $N$ as a
  function of $\Omega$ (see, e.g., Fig~2 of Ref.~\onlinecite{Butts})
  which become negligible at large $\Omega$.
\bibitem{Butts}
D.A. Butts and D.S. Rokhsar,
Nature (London) {\bf 397}, 327 (1999).
\bibitem{commentBucket} In fact, for a vortex in a {\em finite} but
  {\em homogeneous} superfluid (e.g., helium in a ``bucket'', for
  which the superfluid density is finite at the boundary, vanishing only over
  an atomic length scale), the vanishing-current boundary condition is
  crucial for obtaining the correct single-vortex energy and dynamics
  (e.g., the precession of an off-axis vortex). These are strongly
  affected by the interaction of the vortex with its image vortex
  introduced to enforce the above boundary condition.
\bibitem{Inouye}
S. Inouye, S. Gupta, T. Rosenband, A.P. Chikkatur, A. G\'orlitz, T.L. Gustavson,
A.E. Leanhardt, D.E. Pritchard and W. Ketterle,
Phys. Rev. Lett. {\bf 87}, 080402 (2001).
\bibitem{commentFixBC} This is up to a curl-free vector field
  $V(r,\phi)\hat{\phi}$, with $V(r,\phi)$ chosen to match onto the
  inner solution.
\bibitem{comment_farR} For most experimental systems, this regime
  $r\gg R$ is not of physical interest, since the condensate density is
  vanishingly small there (e.g., for the TF $\rho_s(r)$ profile).
  However, this regime may be accessible for condensate profiles with
  spatial variation on multiple length scales, such as $\rho_s(\br)$
  with a Gaussian \lq\lq bump\rq\rq\ on top of a uniform profile
  varying on long length scales, as considered in
  Sec.~\ref{SEC:examples} in the case of {\it many vortices\/} in an
  inhomogeneous condensate.
\bibitem{Ambegaokar}
Vinay Ambegaokar, B.I. Halperin, David R. Nelson and Eric D. Siggia,
Phys. Rev. B {\bf 21}, 1806 (1980).
%
\bibitem{Ao}
See, e.g., Ping Ao and David J. Thouless,
Phys. Rev. Lett., {\bf 70}, 2158 (1993).
%
%
\bibitem{commentTkachenko} For the case of a uniform lattice ($\bu =
  0$), Tkachenko~\cite{Tkachenko} has shown that Eq.~(\ref{eq:tkachenko2}) along with
  Eq.~(\ref{eq:ve}) emerges formally from a small-$z$ expansion of his
  exact expression for  the phase gradient of a uniform vortex array
written in terms of the Weierstrass zeta function
  $\zeta(z)$.
\bibitem{baym1}
Gordon Baym and Elaine Chandler, J. Low Temp. Phys. {\bf 50}, 57 (1983).
\bibitem{Fischer}
Uwe R. Fischer and Gordon Baym,
Phys. Rev. Lett. {\bf 90}, 140402 (2003). 
\bibitem{commentUL} Within our continuum (density-functional)
  approximation the energy functional $E[\bu]$ only depends on the
  vortex density $n_v(r)$ and is therefore independent of the
  transverse component of the phonon field $\bu_T$.
\bibitem{Ho}
Tin-Lun Ho,
Phys. Rev. Lett. {\bf 87}, 060403 (2001).
%
\bibitem{MacDonald}
C.B. Hanna, A.J. Sup, J.C. Diaz-Velez, J. Sinova and A.H. MacDonald,
Bull. Am. Phys. Soc. P28.002 (2004).
\bibitem{Ketterle}
Y. Shin, M. Saba, T.A. Pasquini, W. Ketterle, D.E. Pritchard and A.E. Leanhardt,
Phys. Rev. Lett. {\bf 92}, 050405 (2004).
\bibitem{Tkachenko2}
V.K. Tkachenko,
Zh. Eksp. Teor. Fiz. {\bf 50}, 1573 (1966) 
[Sov. Phys. JETP {\bf 23}, 1049 (1966)];
Zh. Eksp. Teor. Fiz. {\bf 56}, 1763 (1969) 
[Sov. Phys. JETP {\bf 29}, 945 (1969)].
\bibitem{baym03}
Gordon Baym, Phys. Rev. Lett., {\bf 91},  110402 (2003).
\bibitem{Sonin}
E.B. Sonin, cond-mat/0405474.
\bibitem{commentOmega} Quantitative measurement of the rotation rate
  $\Omega$ is in fact quite difficult, as it is typically experimentally
  determined by the cloud aspect ratio with error bars that grow with
  decreasing $\Omega$~\cite{Haljan,Engels}.  
\bibitem{Stauffer}
Dietrich Stauffer and Alexander L. Fetter, Phys. Rev. {\bf 168}, 156 (1968).
\end{thebibliography}
\end{document}